%% file: main.tex
\documentclass[5p, times]{elsarticle}

\usepackage[hyphens]{url}

\usepackage{subcaption}
\usepackage{graphicx}
\usepackage[dvipsnames,svgnames,x11names]{xcolor}
\usepackage{multirow}
\usepackage{booktabs}
\usepackage{tikz,hyperref}
\usepackage{makecell}
\usepackage{longtable}
\usepackage{pdflscape}
\usepackage{hhline}
\usepackage{arydshln}
\usepackage{rotating}
\usepackage{amsmath,amssymb,amsfonts}
\usepackage[linesnumbered,ruled,vlined]{algorithm2e}

\usepackage{algorithmic}
\usepackage{lineno}

\usepackage{textcomp}
\usepackage{csquotes}
\usepackage[nolist,nohyperlinks]{acronym}

\usepackage[final]{changes}

\def\BibTeX{{\rm B\kern-.05em{\sc i\kern-.025em b}\kern-.08em
    T\kern-.1667em\lower.7ex\hbox{E}\kern-.125emX}}

\definecolor{lime}{HTML}{A6CE39}
\DeclareRobustCommand{\orcidicon}{%
    \begin{tikzpicture}
    \draw[lime, fill=lime] (0,0) 
    circle [radius=0.16] 
    node[white] {{\fontfamily{qag}\selectfont \tiny ID}};
    \draw[white, fill=white] (-0.0625,0.095) 
    circle [radius=0.007];
    \end{tikzpicture}
    \hspace{-2mm}
}

\foreach \x in {A, ..., Z}{%
    \expandafter\xdef\csname orcid\x\endcsname{\noexpand\href{https://orcid.org/\csname orcidauthor\x\endcsname}{\noexpand\orcidicon}}
}

\definechangesauthor[color=DarkCyan]{AMS}
\definechangesauthor[color=NavyBlue]{MB}
\definechangesauthor[color=Indigo]{JA}

\begin{document}
\begin{frontmatter}

%


\title{Too Many Options: A Survey of ABE \\ Libraries for Developers}

\author[1,2]{Aintzane Mosteiro-Sanchez}
\ead{amosteiro@ikerlan.es}

\author[1]{Marc Barcelo}
\ead{mbarcelo@ikerlan.es}

\author[2]{Jasone Astorga}
\ead{jasone.astorga@ehu.eus}

\author[1]{Aitor Urbieta}
\ead{AUrbieta@ikerlan.es}

 \affiliation[1]{organization={Ikerlan Technology Research Centre, Basque Research and Technology Alliance (BRTA)},
             addressline={P.J.M. Arizmendiarrieta, 2}, 
             city={Arrasate-Mondragon},
             postcode={20500}, 
             country={Spain}}
            
 \affiliation[2]{organization={University of the Basque Country UPV/EHU},
            addressline={Alameda Urquijo s/n}, 
            city={Bilbao},
            postcode={48013}, 
            country={Spain}}





%

%
\begin{abstract}
Attribute-based encryption (ABE) comprises a set of one-to-many encryption schemes that allow the encryption and decryption of data by associating it with access policies and attributes. Therefore, it is an asymmetric encryption scheme, and its computational requirements limit its deployment in IoT devices. There are different types of ABE and many schemes within each type. However, there is no consensus on the default library for ABE, and those that exist implement different schemes. Developers, therefore, face the challenge of balancing efficiency and security by choosing the suitable library for their projects. This paper studies eleven ABE libraries, analyzing their main features, the mathematical libraries used, and the ABE schemes they provide. The paper also presents an experimental analysis of the four libraries which are still maintained and identifies some of the insecure ABE schemes they implement. In this experimental analysis, we implement the schemes offered by these libraries, measuring their execution times on architectures with different capabilities, i.e., ARMv6 and ARMv8. The experiments provide developers with the necessary information to choose the most suitable library for their projects, according to objective and well-defined criteria.

\end{abstract}


\end{frontmatter}



\input{Docs/acronyms}
\input{Docs/1.Introduction}
\input{Docs/2.Background_State_of_the_art}
\input{Docs/3.Methodology}
\input{Docs/4.Qualitative_Evaluation}
\input{Docs/5.Experimental_evaluation_definition}
\input{Docs/6.Results}

\input{Docs/7.Discussion}

\input{Docs/8.Conclusions}

\input{Docs/ACKs}


\bibliographystyle{elsarticle-num}
\bibliography{references}

\end{document}

%% file: Docs/acronyms.tex
\acrodef{CP-ABE}[CP-ABE]{Ciphertext-Policy Attribute-Based Encryption}
\acrodef{ABE}[ABE]{Attribute-Based Encryption}
\acrodef{dCP-ABE}[dCP-ABE]{Decentralized CP-ABE}
\acrodef{KP-ABE}[KP-ABE]{Key-Policy Attribute-Based Encryption}
\acrodef{PT}[$PT$]{Plaintext}
\acrodef{CT}[$CT$]{Ciphertext}
\acrodef{AP}[$AP$]{Access Policy}
\acrodef{SK}[$SK$]{Secret Key}
\acrodef{DO}[DO]{Data Owner}
\acrodef{RPI4}[RPI4]{Raspberry Pi 4}
\acrodef{RPI0}[RPI0]{Raspberry Pi Zero}
\acrodef{IIoT}[IIoT]{Industrial IoT}
\acrodef{IoT}[IoT]{Internet of Things}
\acrodef{RPI}[RPI]{Raspberry Pi}
\acrodef{IBE}[IBE]{Identity-Based Encryption}

%% file: Docs/1.Introduction.tex
\section{Introduction} \label{Sec:1}
\ac{ABE} is \added{a one-many encryption algorithm. This feature implies that the same piece of encrypted data can be shared with multiple users. Usually, the encryption of a \ac{PT} creates a personalized \ac{CT} which only its unique intended user can decrypt. \ac{ABE} breaks this limitation, improving encryption efficiency for one-to-many communication models. This feature is particularly relevant when the same information must be confidentially shared with multiple users, e.g., brokered communications in industrial environments.}

\replaced{ \ac{ABE} achieves the above feature by providing encryption schemes that bind information encryption and decryption to attributes (e.g., $\mathbb{A} = [att_{1},$ $\ att_{2},\ att_{4},\ att_{5}]$) and an \ac{AP} (e.g., $AP = (att_{1}\ AND\ att_{2})\ OR\ $ $att_{3}$). Since \ac{AP}s are defined according to attributes, access to information is only granted if the policy is fulfilled. As a result, \ac{ABE} provides one-to-many encryption, offers implicit authorization to data, and decouples encryption and decryption from users' identities.}{ABE encryption schemes bind information encryption and decryption to attributes (e.g., $\mathbb{A} = [att_{1},$ $\ att_{2},\ att_{4},\ att_{5}]$) and an \ac{AP} (e.g., $AP = (att_{1}\ AND\ att_{2})\ OR\ $ $att_{3}$). As a result, \ac{ABE} provides one-to-many encryption, offers implicit authorization to data, and decouples encryption and decryption from users' identities.}

\added{Depending on how that fulfillment takes place, two variants of \ac{ABE} have been defined: \ac{KP-ABE} and \ac{CP-ABE}. \ac{KP-ABE} uses \ac{AP}s to create a user's \ac{SK}. Meanwhile, \ac{CT}s are generated according to attributes. If the attributes of the \ac{CT} fulfill the \ac{AP} of the user, the user gets access to information.} \ac{CP-ABE} does the opposite of \ac{KP-ABE}: users receive \ac{SK}s generated according to attributes, and information is protected according to \ac{AP}. In both cases, attributes must satisfy the \ac{AP} to decrypt the information. 

\added{Although the concept of fulfilling \ac{AP} in order to access information is similar in both \ac{ABE} schemes, it can be seen how \ac{CP-ABE} allows the \ac{DO} to retain control over who can access the information. In distributed environments, it is more secure that whoever generates the information is the one who decides the access policy to it. Therefore \ac{CP-ABE} is more suitable for these environments. Besides, defining users according to attributes makes the system more scalable since it is more manageable to update users' privileges by adding or deleting attributes rather than redefining their \ac{AP} by keeping a list of every resource they need to access.}

\replaced{There are numerous \ac{KP-ABE} and \ac{CP-ABE} schemes, and not all of them are implemented in every library. Therefore, knowing which libraries are more efficient or secure is a challenge.}{There are numerous \ac{KP-ABE} and \ac{CP-ABE} schemes, and libraries do not always implement the same ones. Thus, it is challenging to know which libraries are more efficient or secure.} Choosing the proper \ac{ABE} library lies in a trade-off between efficiency and security, which can be a challenge: Which \ac{ABE} schemes do they offer? Which programming languages are available? Which mathematical libraries do they use? In order to provide answers to these questions, this article surveys existing cryptographic libraries and provides a fair comparison according to their security features \added{and performance}. \replaced{The goal is to assist developers in choosing the most suitable library for their projects. This article is an extended version of a preliminary work presented in~\cite{mosteiroURSI}. The contributions of this article can be summarized as:}{The goal is to assist developers in choosing the most appropriate library for their projects.}

\begin{itemize}
    \item \added{An experimental evaluation of \ac{ABE} libraries, which is a practical tool to help developers choose the most suitable one for their projects.}
    \item \added{The methodology itself, which can be further used to evaluate new ABE schemes, increasing the scope of the assessment.}
\end{itemize}

%% file: Docs/2.Background_State_of_the_art.tex
\section{Background \& Related Work} \label{Sec:2}

\added{The first \ac{ABE} algorithm was presented under the name of Fuzzy Identity-Based Encryption~\cite{sahai2005fuzzy}, giving more granularity to \ac{IBE} schemes~\cite{10.1007/3-540-39568-7_5}. \ac{IBE} schemes encrypt and decrypt information according to the receivers' identities. The scheme presented in~\cite{sahai2005fuzzy} provided more flexibility to encryption by using attributes instead of identities. Fuzzy Identity-Based Encryption eventually led to the term \ac{ABE} and the development of a new family of asymmetric encryption algorithms. \ac{ABE} continued to be developed in the cryptographic field, giving rise to \ac{KP-ABE} \cite{10.1145/1180405.1180418} and \ac{CP-ABE} \cite{4223236}; and even a decentralized version called \ac{dCP-ABE} \cite{10.1007/978-3-642-00730-9_2}.}

\begin{figure*}[!hbtp]
    \centerline{\includegraphics[width=0.7\textwidth]{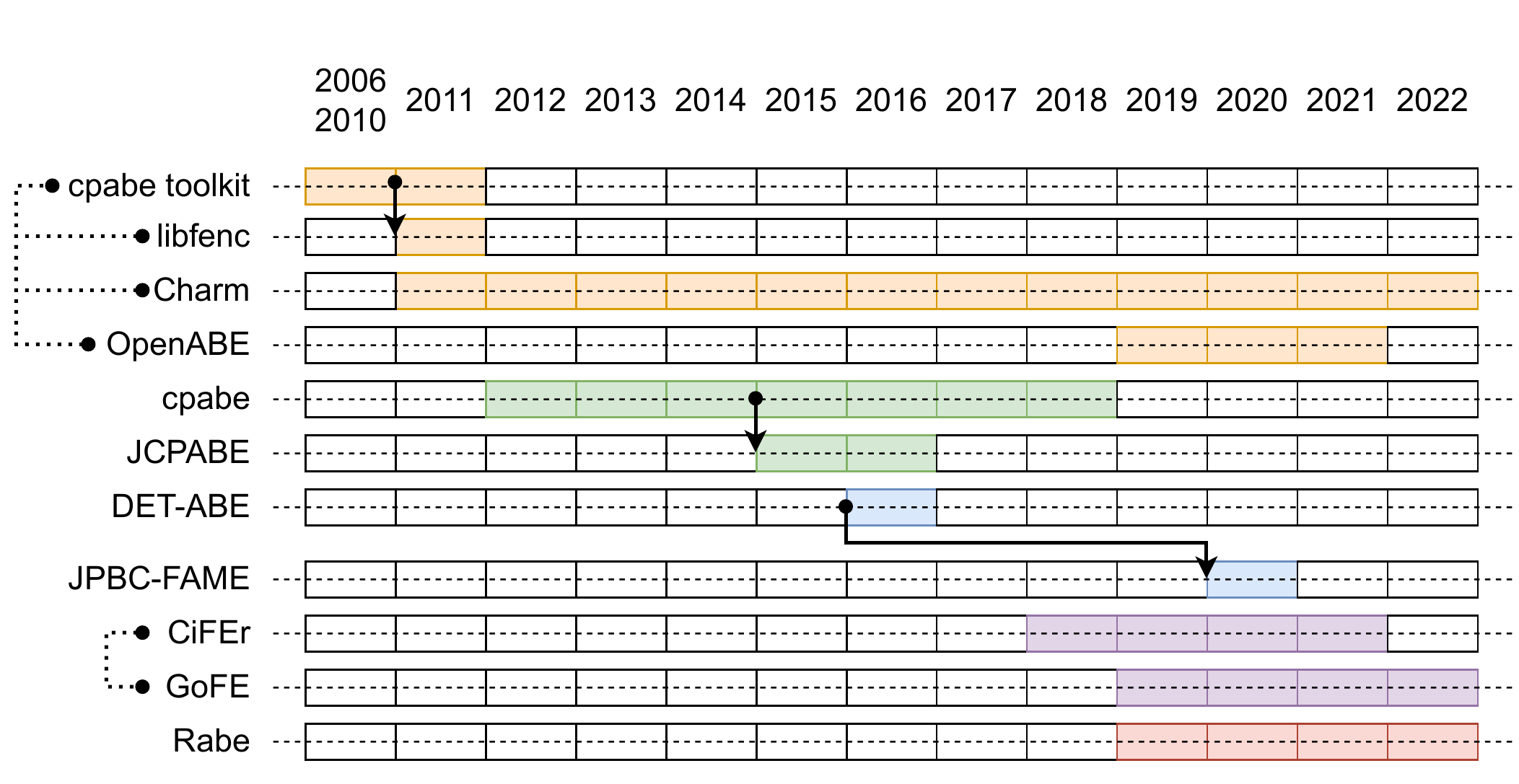}}
    \caption{Library timeline. Dashed lines represent libraries that share developers. Continuous line arrows represent libraries that reuse code. Colors show libraries with some relationship between them (code or developers).}
    \label{fig:FIG0}
\end{figure*}

\added{However, for a cryptographic scheme to develop beyond academia, practical implementations should be available in the form of libraries. Developers use cryptography as a tool, so part of the popularity of certain schemes is based on the ease of obtaining a library. When Bethencourt \textit{et al.} presented CP-ABE~\cite{4223236}, they included the \textit{cpabe toolkit}~\cite{cpabe-toolkit}. This library was, until 2011, the only library offering \ac{ABE} schemes. However, in 2011 \textit{libfenc}~\cite{libfenc} appeared. This project reused part of the \textit{cpabe toolkit} code but extended the functions and schemes of the library. These relationships are shown in Figure \ref{fig:FIG0} by solid line arrows.}

\added{The same year \textit{libfenc} appeared, the first version of \textit{Charm}~\cite{Charm} was released. This library focuses on elliptic curve cryptography. Several \textit{libfenc} developers have also participated in the development and maintenance of \textit{Charm}, which, as of May 2022, is still being maintained and updated. On the other hand, it is also interesting to mention that some of the \textit{cpabe toolkit} and \textit{Charm} developers worked together and released a new library for \ac{ABE} in 2019, called \textit{OpenABE}~\cite{OpenABE}. This relationship of developers is shown in Figure \ref{fig:FIG0} by dashed lines next to the name of the libraries.}

\added{In general, there are more \ac{ABE} schemes in Academia than libraries capable of implementing them because not every proposal includes a library. One of the exceptions is the case of BDABE~\cite{secrypt18}, developed by Fraunhofer, which was accompanied by \textit{Rabe}~\cite{Rabe}. Moreover, \textit{Rabe} implements several other \ac{ABE} schemes, not only BDABE, and it is still maintained. Despite these advances, there is still a need for a production-grade library. Therefore, part of the results of the FENTEC}\footnote{https://fentec.eu/} \added{ project, which aimed to develop functional encryption systems, resulted in two \ac{ABE} libraries, \textit{GoFE}~\cite{GoFE} and \textit{CiFEr}~\cite{CiFEr}, between 2018 and 2019. Both implement the same ABE schemes, but \textit{GoFE} is implemented in Go and \textit{CiFEr} in C.}


\added{To deploy \ac{ABE} encryption schemes properly, their efficiency and complexity cannot have a negative impact on the system.} 
Authors in~\cite{cryptoeprint:2022:038} study the \added{mathematical} complexity of various \ac{CP-ABE} schemes and propose different strategies to optimize them. They conclude that there is no \added{generic} solution capable of reducing the complexity \added{of \ac{ABE} schemes} in every step. Instead, users must prioritize reducing the complexity of encryption, decryption, or key generation. \added{Depending on the system requirements, authors suggest several mathematical improvements on the different \ac{ABE} schemes.} 

\deleted{Interested readers are referred to the above source for guidance on these complexity considerations, as we consider it out-of-scope for the present article. Instead, we follow a  more practical approach and focus on studying the execution time of \ac{ABE} libraries, what parameters should be considered for their choice, and which libraries should be discarded.}

\replaced{Although mathematical efficiency is crucial for \ac{ABE} schemes, their practical implementation is not always optimal. Dependencies on third parties, implementation errors, or programming language affect execution times. Furthermore, mathematical dependencies also affect the security of the implementations. Therefore, formal analyses of cryptographic schemes are insufficient, and a more practical approach is required.}{Thus, we consider a library's execution time and security crucial features to select it. Time can be measured by implementing the libraries, but security evaluations are more challenging. For this paper, we base part of our analysis on the results of~\cite{10.1007/978-3-030-75539-3_5}. In the cited article, researchers analyze many existing \ac{ABE} schemes, study their vulnerabilities, propose new attacks, and provide an extensive list of vulnerable schemes.}

\added{A library's execution time and security are crucial features to select it. Time can be measured by implementing the libraries, but security evaluations are more challenging. For this article, part of the analysis is based on the results of~\cite{10.1007/978-3-030-75539-3_5}. In the cited article, the researchers analyze several ABE schemes, study their vulnerabilities, propose new attacks, and provide an extensive list of vulnerable schemes.} 

%% file: Docs/3.Methodology.tex
\section{Methodology} \label{Sec:3}
\begin{figure}[!ht]
\centerline{\includegraphics[width=0.7\columnwidth]{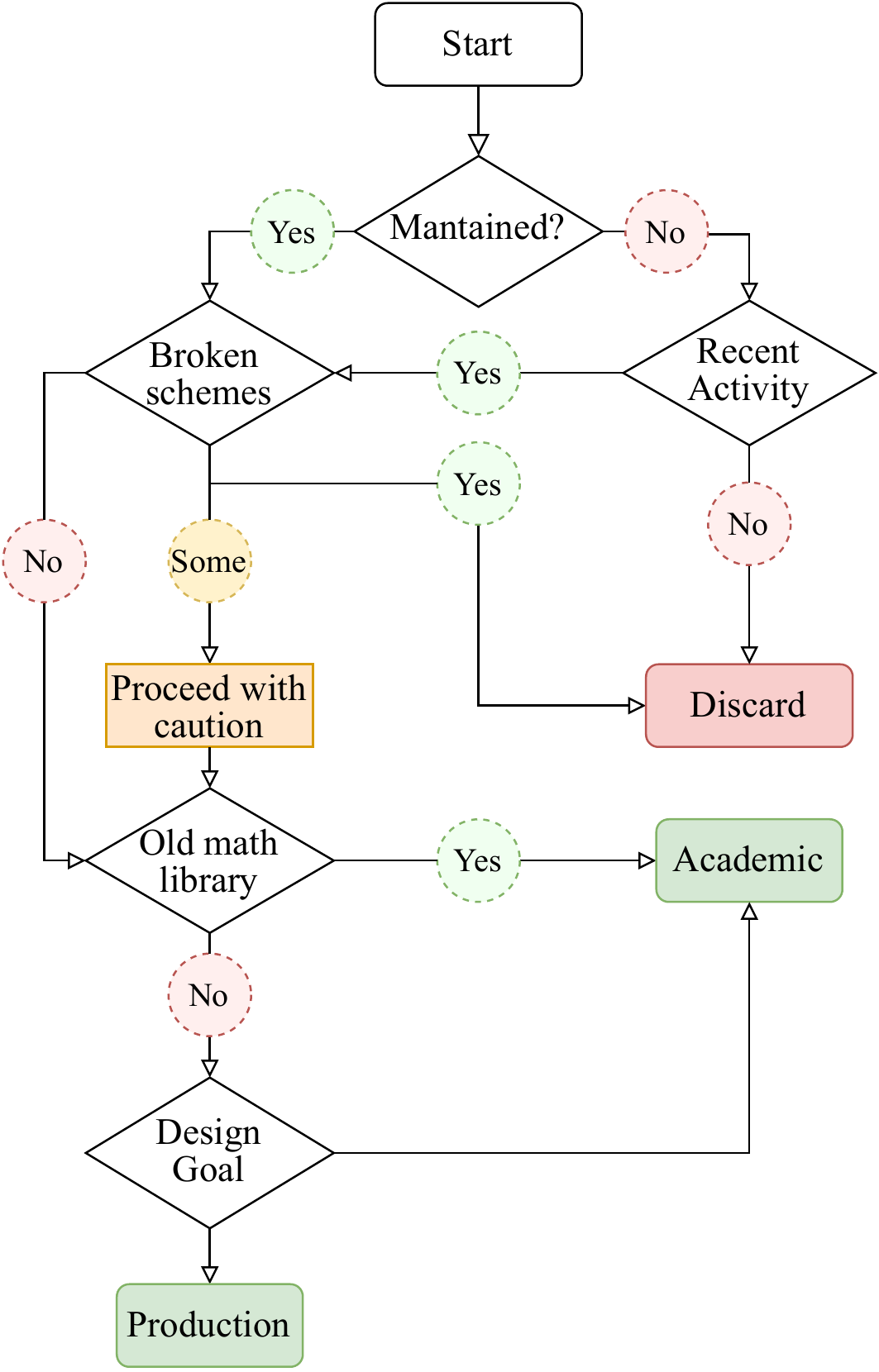}}\caption{Library analysis process}
\label{fig:FIG1}
\end{figure}

We provide two evaluations for \ac{ABE} libraries: qualitative and experimental. Every library is qualitatively evaluated, but some are dismissed after the assessment and are not experimentally analyzed. This section defines which features are considered and describes the evaluation process \added{(Figure \ref{fig:FIG1})}.

\begin{itemize}
    \item \textbf{Maintained:} We consider a library to be maintained when its main fork has had some \replaced{update or upgrade}{activity} in the last fourteen months \added{as of may 2022}.
    \item \textbf{Recent Activity:} We consider that there is recent activity when the main fork of the library has been \replaced{maintained}{updated in the last fourteen months \added{as of may 2022}} or if any branch has had activity in the last year.
    \item \textbf{Broken Schemes:} \replaced{Some ABE schemes have been successfully attacked.}{The are many ABE schemes, and some have been successfully attacked.} The authors of~\cite{10.1007/978-3-030-75539-3_5} compile a list of 24 vulnerable ABE schemes. Interested readers are encouraged to consult Tables 1, 2, and 5 of the mentioned article for a more detailed analysis.
    \item \textbf{Math Library:} ABE schemes \added{are based on elliptic curve cryptography, but} do not require specific curves; instead, developers are the ones to choose them. Thus, the chosen mathematical library is crucial for ABE security and efficiency. Some mathematical libraries support outdated curves and should only be used in academic or experimental environments.
    \item \textbf{Design \replaced{Purpose}{goal}:} Not every library is intended or appropriate for production. Understanding the use of each library is critical for deploying a secure system. \added{Libraries not designed for production environments should not be considered in such scenarios.}
\end{itemize}

All the libraries are studied according to these five features. Then, following the process shown in Figure \ref{fig:FIG1}, we choose the libraries to be experimentally analyzed. We consider that libraries implementing outdated curves are unsuitable for the production environment, regardless of their design goal.


%% file: Docs/4.Qualitative_Evaluation.tex
\section{Qualitative Evaluation} \label{Sec:4}

\subsection{Generic Features} \label{Sec:4.1}
\replaced{This section uses the criteria defined in the previous section to evaluate a total of 11 libraries.}{In this section, the selected libraries are evaluated according to the criteria defined in the previous section.} \replaced{The information gathered about this libraries}{The gathered information} is presented in Table \ref{tab:1}, which is analyzed according to the process shown in Figure \ref{fig:FIG1}. \replaced{Information on whether there is an active community using the libraries has been included in the table. This is not such a crucial criterion as to merit inclusion in the selection process in Figure \ref{fig:FIG1}, but it may \replaced{be of interest to}{interest} developers. Note that an active community can provide technical support, even if the library is no longer maintained.}{Information about active community has been included since developers can have technical support from the community even if the library has been discontinued.}

\begin{table*}[!htb]
\renewcommand{\arraystretch}{1.1}
\caption{Comparison of libraries. Part I} \label{tab:1}
\begin{center}
\resizebox{0.76\textwidth}{!}{
\begin{tabular}{lllllll}
\hline \hline
Name

& \begin{tabular}[c]{@{}l@{}} Mantained \\ \fbox{\tiny{1}} \fbox{\tiny{2}} \end{tabular}
& \begin{tabular}[c]{@{}l@{}} Active \\Community \\ \fbox{\tiny{1}} \fbox{\tiny{2}} \end{tabular}
& \begin{tabular}[c]{@{}l@{}} Recent \\Activity \\ \fbox{\tiny{1}} \fbox{\tiny{2}}\end{tabular}
& \begin{tabular}[c]{@{}l@{}} Broken \\Schemes \\ \fbox{\tiny{1}} \fbox{\tiny{2}} \fbox{\tiny{3}} \end{tabular}
& \begin{tabular}[c]{@{}l@{}} Math Library \\ \fbox{\tiny{1}} \fbox{\tiny{2}} \fbox{\tiny{3}} \fbox{\tiny{4}} \fbox{\tiny{5}} \fbox{\tiny{6}} \fbox{\tiny{7}} \fbox{\tiny{8}} \fbox{\tiny{9}} \end{tabular} 
& \begin{tabular}[c]{@{}l@{}} Design \\ Goal \\ \fbox{\tiny{1}} \fbox{\tiny{2}} \end{tabular} \\

\hline \hline
& \begin{tabular}[c]{@{}l@{}} 1. Yes \\ 2. No \end{tabular} & \begin{tabular}[c]{@{}l@{}} 1. Yes \\ 2. No\end{tabular} & \begin{tabular}[c]{@{}l@{}} 1. Yes \\ 2. No\end{tabular} & \begin{tabular}[c]{@{}l@{}} 1. Yes \\ 2. No\\ 3. Some\end{tabular} 

& \begin{tabular}{ll} 
1. Mircl        & 6. Relic \\
2. libsodium    & 7. OpenSSL \\
3. bn-256       & 8. jPBC \\
4. go-crypto    & 9. PBC \\
5. Rabe-bn       
\end{tabular} 

& \begin{tabular}[c]{@{}l@{}} 1. Production \\ 2. Research \end{tabular} \\

\hline \hline

libfenc~\cite{libfenc}
& $\square$\hspace{6.5pt}$\blacksquare$
& $\square$\hspace{6.5pt}$\blacksquare$ 
& $\square$\hspace{6.5pt}$\blacksquare$       
& $\square$\hspace{6.5pt}$\blacksquare$\hspace{6.5pt}$\square$                 
& $\square$\hspace{6.5pt}$\square$\hspace{6.5pt}$\square$\hspace{6.5pt}$\square$\hspace{6.5pt}$\square$\hspace{6.5pt}$\square$\hspace{6.5pt}$\square$\hspace{6.5pt}$\square$\hspace{6.5pt}$\blacksquare$ 
& $\square$\hspace{6.5pt}$\blacksquare$\\
\hline

cpabe-toolkit~\cite{cpabe-toolkit}
& $\square$\hspace{6.5pt}$\blacksquare$
& $\square$\hspace{6.5pt}$\blacksquare$ 
& $\square$\hspace{6.5pt}$\blacksquare$       
& $\square$\hspace{6.5pt}$\blacksquare$\hspace{6.5pt}$\square$                 
& $\square$\hspace{6.5pt}$\square$\hspace{6.5pt}$\square$\hspace{6.5pt}$\square$\hspace{6.5pt}$\square$\hspace{6.5pt}$\square$\hspace{6.5pt}$\square$\hspace{6.5pt}$\square$\hspace{6.5pt}$\blacksquare$ 
& $\square$\hspace{6.5pt}$\blacksquare$\\
\hline

\textbf{OpenABE~\cite{OpenABE}}
& $\square$\hspace{6.5pt}$\blacksquare$
& $\blacksquare$\hspace{6.5pt}$\square$ 
& $\blacksquare$\hspace{6.5pt}$\square$       
& $\square$\hspace{6.5pt}$\blacksquare$\hspace{6.5pt}$\square$                 
& $\square$\hspace{6.5pt}$\square$\hspace{6.5pt}$\square$\hspace{6.5pt}$\square$\hspace{6.5pt}$\square$\hspace{6.5pt}$\blacksquare$\hspace{6.5pt}$\blacksquare$\hspace{6.5pt}$\square$\hspace{6.5pt}$\square$ 
& $\blacksquare$\hspace{6.5pt}$\square$\\
\hline

JPBC-FAME~\cite{JPBC-FAME}
& $\square$\hspace{6.5pt}$\blacksquare$
& $\square$\hspace{6.5pt}$\blacksquare$ 
& $\square$\hspace{6.5pt}$\blacksquare$       
& $\square$\hspace{6.5pt}$\blacksquare$\hspace{6.5pt}$\square$                 
& $\square$\hspace{6.5pt}$\square$\hspace{6.5pt}$\square$\hspace{6.5pt}$\square$\hspace{6.5pt}$\square$\hspace{6.5pt}$\square$\hspace{6.5pt}$\square$\hspace{6.5pt}$\blacksquare$\hspace{6.5pt}$\square$ 
& $\square$\hspace{6.5pt}$\blacksquare$\\
\hline

cp-abe~\cite{cp-abe}
& $\square$\hspace{6.5pt}$\blacksquare$
& $\square$\hspace{6.5pt}$\blacksquare$ 
& $\square$\hspace{6.5pt}$\blacksquare$       
& $\square$\hspace{6.5pt}$\blacksquare$\hspace{6.5pt}$\square$                 
& $\square$\hspace{6.5pt}$\square$\hspace{6.5pt}$\square$\hspace{6.5pt}$\square$\hspace{6.5pt}$\square$\hspace{6.5pt}$\square$\hspace{6.5pt}$\square$\hspace{6.5pt}$\blacksquare$\hspace{6.5pt}$\square$ 
& $\square$\hspace{6.5pt}$\blacksquare$\\
\hline

DET-ABE~\cite{DET-ABE}
& $\square$\hspace{6.5pt}$\blacksquare$
& $\square$\hspace{6.5pt}$\blacksquare$ 
& $\square$\hspace{6.5pt}$\blacksquare$       
& $\square$\hspace{6.5pt}$\blacksquare$\hspace{6.5pt}$\square$                 
& $\square$\hspace{6.5pt}$\square$\hspace{6.5pt}$\square$\hspace{6.5pt}$\square$\hspace{6.5pt}$\square$\hspace{6.5pt}$\square$\hspace{6.5pt}$\square$\hspace{6.5pt}$\blacksquare$\hspace{6.5pt}$\square$ 
& $\square$\hspace{6.5pt}$\blacksquare$\\
\hline

\textbf{Rabe~\cite{Rabe}}
& $\blacksquare$\hspace{6.5pt}$\square$
& $\blacksquare$\hspace{6.5pt}$\square$ 
& $\blacksquare$\hspace{6.5pt}$\square$       
& $\square$\hspace{6.5pt}$\square$\hspace{6.5pt}$\blacksquare$                 
& $\square$\hspace{6.5pt}$\square$\hspace{6.5pt}$\square$\hspace{6.5pt}$\square$\hspace{6.5pt}$\blacksquare$\hspace{6.5pt}$\square$\hspace{6.5pt}$\square$\hspace{6.5pt}$\square$\hspace{6.5pt}$\square$ 
& $\blacksquare$\hspace{6.5pt}$\square$\\
\hline

JCPABE~\cite{JCPABE}
& $\square$\hspace{6.5pt}$\blacksquare$
& $\square$\hspace{6.5pt}$\blacksquare$ 
& $\square$\hspace{6.5pt}$\blacksquare$       
& $\square$\hspace{6.5pt}$\blacksquare$\hspace{6.5pt}$\square$                 
& $\square$\hspace{6.5pt}$\square$\hspace{6.5pt}$\square$\hspace{6.5pt}$\square$\hspace{6.5pt}$\square$\hspace{6.5pt}$\square$\hspace{6.5pt}$\square$\hspace{6.5pt}$\blacksquare$\hspace{6.5pt}$\square$
& $\square$\hspace{6.5pt}$\blacksquare$\\
\hline

\textbf{Charm~\cite{Charm}}
& $\blacksquare$\hspace{6.5pt}$\square$
& $\blacksquare$\hspace{6.5pt}$\square$ 
& $\blacksquare$\hspace{6.5pt}$\square$       
& $\square$\hspace{6.5pt}$\square$\hspace{6.5pt}$\blacksquare$                 
& $\blacksquare$\hspace{6.5pt}$\square$\hspace{6.5pt}$\square$\hspace{6.5pt}$\square$\hspace{6.5pt}$\square$\hspace{6.5pt}$\blacksquare$\hspace{6.5pt}$\blacksquare$\hspace{6.5pt}$\square$\hspace{6.5pt}$\blacksquare$ 
& $\blacksquare$\hspace{6.5pt}$\square$\\
\hline

\textbf{GoFE~\cite{GoFE}}
& $\blacksquare$\hspace{6.5pt}$\square$
& $\blacksquare$\hspace{6.5pt}$\square$ 
& $\blacksquare$\hspace{6.5pt}$\square$       
& $\square$\hspace{6.5pt}$\blacksquare$\hspace{6.5pt}$\square$                 
& $\square$\hspace{6.5pt}$\square$\hspace{6.5pt}$\blacksquare$\hspace{6.5pt}$\blacksquare$\hspace{6.5pt}$\square$\hspace{6.5pt}$\square$\hspace{6.5pt}$\square$\hspace{6.5pt}$\square$\hspace{6.5pt}$\square$ 
& $\square$\hspace{6.5pt}$\blacksquare$\\
\hline

CiFEr~\cite{CiFEr}
& $\square$\hspace{6.5pt}$\blacksquare$
& $\blacksquare$\hspace{6.5pt}$\square$ 
& $\blacksquare$\hspace{6.5pt}$\square$       
& $\square$\hspace{6.5pt}$\blacksquare$\hspace{6.5pt}$\square$                 
& $\blacksquare$\hspace{6.5pt}$\blacksquare$\hspace{6.5pt}$\square$\hspace{6.5pt}$\square$\hspace{6.5pt}$\square$\hspace{6.5pt}$\square$\hspace{6.5pt}$\square$\hspace{6.5pt}$\square$\hspace{6.5pt}$\square$ 
& $\square$\hspace{6.5pt}$\blacksquare$\\
\hline \hline

\end{tabular}
}
\end{center}
\end{table*}

Table \ref{tab:1} shows that only \added{three libraries continue to be maintained by the original developers in the main fork:} \textit{Rabe}, \textit{Charm} and \textit{GoFE}. \added{Of the remainder, it should be noted that most had no activity in the main fork in almost two years. The exceptions to this are \textit{OpenABE} and \textit{CiFEr}. \textit{OpenABE} was last updated in the main fork on January 2021 and \textit{CiFEr} on February 2021. However, both of them have had recent activity in new forks. In the case of \textit{OpenABE} fork}\footnote{https://github.com/StefanoBerlato/openabe}, \added{it solves installation issues and updates dependencies like OpenSSL. In the case of the new \textit{CiFEr} fork}\footnote{https://github.com/swanhong/CiFEr/tree/master}, \added{ it has received minor updates. It should also be noted that \textit{CiFEr} is the twin of \textit{GoFE} but implemented in C instead of Golang. In other words, although the main branch of \textit{CiFEr} has no activity, the project is still going on. Therefore, this article considers that all five libraries, i.e., \textit{OpenABE}, \textit{Rabe}, \textit{Charm}, \textit{GoFE}, and \textit{CiFEr}, have had recent activity.} \deleted{We define recent activity as any fork derived from the main fork being maintained, even if it does not have an active community supporting it. }

\added{Regarding community activity, \textit{Rabe} and \textit{Charm} are the most active, although in different ways. \textit{Rabe} developers respond and resolve open issues in the repository and update the library with user-requested features}\footnote{https://github.com/Fraunhofer-AISEC/rabe/issues/9}. \added{\textit{Charm} is the most widely used framework for cryptographic prototyping, and as such, it has an extensive community of users.} \replaced{An active community favors problem resolution, which makes it an interesting feature to be considered. However, it is not so critical that its absence implies discarding the library.}{An active community guarantees problem solving; thus, it is interesting to consider it, although it is not a critical feature to dismiss its implementation.}

\subsection{Security Features} \label{Sec:4.2}

\deleted{In terms of security, \textit{Charm} and \textit{Rabe} support the \ac{KP-ABE} scheme proposed in~\cite{YAO2015104}. In addition, \textit{Charm} also implements DAC-MACs~\cite{Yang2014}, also broken in~\cite{10.1007/978-3-030-75539-3_5}. Broken schemes are insecure and should never be used. However, both \textit{Charm} and \textit{Rabe} support other \ac{ABE} schemes, so there is no need to discard them. Instead, developers should use these libraries with caution and warrant that vulnerable schemes are not implemented.}

\added{In this section, we analyze those features directly affecting the libraries' security: the chosen mathematical libraries, the implemented ABE schemes, and the AES schemes underneath.}

\replaced{Regarding the mathematical libraries used, as Table \ref{tab:1} showed,}{As for the mathematical library,} some libraries use PBC, which supports obsolete elliptic curves that currently are not considered secure enough for production~\cite{cryptoeprint:2022:038}. Therefore, we consider libraries using PBC only suitable for research. Java libraries use jPBC, the Java implementation of the original PBC, and therefore implements the same elliptic curves as PBC. \replaced{Libraries like \textit{Charm} use PBC but can be compiled with Mircl or Relic. Relic, PBC, and Mircl are pairing-based cryptographic libraries. However, Relic offers more efficient pairing constructions and faster implementations than PBC, and Mircl is newer and available in seven programming languages. Mircl is also the mathematical library used by \textit{CiFEr}, albeit with a crucial nuance compared to \textit{Charm}:}{Some libraries like \textit{Charm} use PBC by default but can be compiled with Mircl or Relic. Another library using Mircl is \textit{CiFEr}. However, there is an important nuance:} \textit{Charm} uses the maintained branch of Mircl. In contrast, \textit{CiFEr} uses Mircl-AMCL, a non-maintained branch of Mircl. Moreover, \textit{CiFEr} implements a reduced version of Mircl-AMCL, using the BN-254 curve for 64 bits. \added{As a result, \textit{CiFEr} can only be implemented in a 64-bit architecture, limiting its deployability.}  In contrast, \textit{Charm} imposes no restrictions on the curve to use from Mircl-Core. Developers should take this into account and modify the files as necessary. \added{With the architecture limitation of \textit{CiFEr}, \textit{OpenABE} becomes the only library written in C++ with no architecture limitation. A C++ library is especially relevant for \mbox{\ac{IoT}} devices with minimal computational power. This, as well as some embedded devices, benefit from the use of C and C++.}

\replaced{Regarding ``Design Goal,"}{Finally, when we mention ``Design Goal,"} we refer to research or production quality. We consider cryptographic libraries using obsolete curves suitable for research and academic purposes but not production. Furthermore, some developers flag their libraries as unsuitable for production, either because they have not been adequately tested or are still in an early stage of development. Thus, \textit{OpenABE}, \textit{Rabe}, and \textit{Charm} are the libraries currently suitable for production \added{while \textit{GoFE} should still be limited to academic and research environments}. However, as pointed out earlier, \textit{Charm} should be compiled with Mircl or Relic, not PBC. \added{With the four main libraries identified, we examine their implementation in Table \ref{tab:2}. To enhance results' readability, these four libraries have also been highlighted in bold in Table \ref{tab:1}.}\deleted{To enhance results' readability, these three libraries have been highlighted in bold in Table \ref{tab:1}.}

\begin{table*}[!htb]
\renewcommand{\arraystretch}{1.1}
\caption{Comparison of libraries. Part II} \label{tab:2}
\begin{center}
\resizebox{0.75\textwidth}{!}{
\begin{tabular}{lllllll}
\hline \hline
Name
& \begin{tabular}[c]{@{}l@{}} Docs. \\ \fbox{\tiny{1}} \fbox{\tiny{2}}\end{tabular}
& \begin{tabular}[c]{@{}l@{}} Language \\ \fbox{\tiny{1}} \fbox{\tiny{2}} \fbox{\tiny{3}} \fbox{\tiny{4}} \fbox{\tiny{5}} \end{tabular}
& \begin{tabular}[c]{@{}l@{}} KP-ABE \\ \fbox{\tiny{1}} \fbox{\tiny{2}} \fbox{\tiny{3}} \fbox{\tiny{4}} \end{tabular}
& \begin{tabular}[c]{@{}l@{}} CP-ABE \\ \fbox{\tiny{1}} \fbox{\tiny{2}} \fbox{\tiny{3}} \fbox{\tiny{4}} \fbox{\tiny{5}} \fbox{\tiny{6}} \end{tabular}
& \begin{tabular}[c]{@{}l@{}} dCP-ABE \\ \fbox{\tiny{1}} \fbox{\tiny{2}} \fbox{\tiny{3}} \fbox{\tiny{4}} \fbox{\tiny{5}} \fbox{\tiny{6}}\end{tabular} \\

\hline \hline

& \begin{tabular}[c]{@{}l@{}} 1. Yes \\ 2. No \end{tabular}


& \begin{tabular}[c]{@{}l@{}} 1. C++ \\ 2. Rust \\ 3. Go \\ 4. Python \end{tabular} 

& \begin{tabular}[c]{@{}l@{}} 1. GPSW06~\cite{10.1145/1180405.1180418} \\ 2. LSW10~\cite{5504791}\\ 3. YCT14~\cite{YAO2015104} \textasteriskcentered  \\ 4. FAME~\cite{10.1145/3133956.3134014} \end{tabular} 

& \begin{tabular}[c]{@{}l@{}} 1. BSW07~\cite{4223236} \\ 2. W11~\cite{waters2011ciphertext} \\ 3. YAHK14~\cite{10.1007/978-3-642-54631-0_16} \\ 4. CGW15~\cite{10.1007/978-3-662-46803-6_20} \\ 5. TimePRE~\cite{LIU2014355} \\ 6. FAME~\cite{10.1145/3133956.3134014} \end{tabular} 


& \begin{tabular}[c]{@{}l@{}} 1. MKE08~\cite{10.1007/978-3-642-00730-9_2} \\ 2. LW11~\cite{lewko2011decentralizin}\\ 3. DAC-MACS~\cite{Yang2014} \textasteriskcentered  \\ 4. YJ14~\cite{6620875} \textasteriskcentered  \\ 5. RW15~\cite{10.1007/978-3-662-47854-7_19} \\ 6. BDABE~\cite{secrypt18} \end{tabular} \\


\hline \hline

OpenABE~\cite{OpenABE}
& $\blacksquare$\hspace{6.5pt}$\square$
& $\blacksquare$\hspace{6.5pt}$\square$\hspace{6.5pt}$\square$\hspace{6.5pt}$\square$
& $\blacksquare$\hspace{6.5pt}$\square$\hspace{6.5pt}$\square$\hspace{6.5pt}$\square$ 
& $\square$\hspace{6.5pt}$\blacksquare$\hspace{6.5pt}$\square$\hspace{6.5pt}$\square$\hspace{6.5pt}$\square$\hspace{6.5pt}$\square$
& $\square$\hspace{6.5pt}$\square$\hspace{6.5pt}$\square$\hspace{6.5pt}$\square$\hspace{6.5pt}$\square$\hspace{6.5pt}$\square$ \\
\hline
 
   Rabe~\cite{Rabe}
& $\blacksquare$\hspace{6.5pt}$\square$
& $\square$\hspace{6.5pt}$\blacksquare$\hspace{6.5pt}$\square$\hspace{6.5pt}$\square$
& $\square$\hspace{6.5pt}$\blacksquare$\hspace{6.5pt}$\blacksquare$\hspace{6.5pt}$\blacksquare$ 
& $\blacksquare$\hspace{6.5pt}$\square$\hspace{6.5pt}$\square$\hspace{6.5pt}$\square$\hspace{6.5pt}$\square$\hspace{6.5pt}$\blacksquare$
& $\blacksquare$\hspace{6.5pt}$\blacksquare$\hspace{6.5pt}$\square$\hspace{6.5pt}$\square$\hspace{6.5pt}$\square$\hspace{6.5pt}$\blacksquare$ \\
\hline

    Charm~\cite{Charm}
& $\blacksquare$\hspace{6.5pt}$\square$
& $\square$\hspace{6.5pt}$\square$\hspace{6.5pt}$\square$\hspace{6.5pt}$\blacksquare$
& $\square$\hspace{6.5pt}$\blacksquare$\hspace{6.5pt}$\blacksquare$\hspace{6.5pt}$\square$ 
& $\blacksquare$\hspace{6.5pt}$\blacksquare$\hspace{6.5pt}$\blacksquare$\hspace{6.5pt}$\blacksquare$\hspace{6.5pt}$\blacksquare$\hspace{6.5pt}$\blacksquare$
& $\square$\hspace{6.5pt}$\blacksquare$\hspace{6.5pt}$\blacksquare$\hspace{6.5pt}$\blacksquare$\hspace{6.5pt}$\blacksquare$\hspace{6.5pt}$\square$ \\
\hline

GoFE~\cite{GoFE}
& $\blacksquare$\hspace{6.5pt}$\square$
& $\square$\hspace{6.5pt}$\square$\hspace{6.5pt}$\blacksquare$\hspace{6.5pt}$\square$
& $\blacksquare$\hspace{6.5pt}$\square$\hspace{6.5pt}$\square$\hspace{6.5pt}$\square$ 
& $\square$\hspace{6.5pt}$\square$\hspace{6.5pt}$\square$\hspace{6.5pt}$\square$\hspace{6.5pt}$\square$\hspace{6.5pt}$\blacksquare$
& $\square$\hspace{6.5pt}$\blacksquare$\hspace{6.5pt}$\square$\hspace{6.5pt}$\square$\hspace{6.5pt}$\square$\hspace{6.5pt}$\square$\\
\hline \hline 

\multicolumn{1}{l}{\textasteriskcentered Broken scheme}
\end{tabular}
}
\end{center}
\end{table*}

\added{In terms of supported schemes, \textit{Charm} and \textit{Rabe} support YCT14, a \ac{KP-ABE} scheme proposed in~\cite{YAO2015104}, which was broken in 2019~\cite{8651482}. It is also one of the broken schemes compiled in~\cite{10.1007/978-3-030-75539-3_5}. In addition, \textit{Charm} also implements some vulnerable \ac{dCP-ABE} schemes: YJ14~\cite{6620875} and DAC-MACs~\cite{Yang2014}, both broken in~\cite{10.1007/978-3-030-75539-3_5}. Broken schemes are insecure and should never be used. However, both \textit{Charm} and \textit{Rabe} support other \ac{ABE} schemes, so there is no need to discard them. Instead, developers should use these libraries cautiously and warrant that vulnerable schemes are not implemented. Another scheme that should be carefully considered is BDABE~\cite{secrypt18}. This is the only \ac{ABE} scheme designed to work with Blockchain. Thus, as the authors of the scheme explain, developers should consider that the scheme's efficiency will be affected by the deployed Blockchain solution.}

\added{As mentioned earlier, ABE schemes are computationally heavy and are often used in hybrid mode. Hybrid mode consists of using a symmetric scheme (usually AES) to encrypt the message and an ABE scheme to encrypt the symmetric key. Therefore, to properly analyze the libraries, we also study the implemented AES schemes. Table \ref{tab:3} shows the result of this evaluation.}

\begin{table}[!htb]
\renewcommand{\arraystretch}{1}
\caption [Implemented symmetric ciphers]{\label{tab:3} Implemented symmetric ciphers.}
\begin{center}
\resizebox{0.73\columnwidth}{!}{
\begin{tabular}{lcc}
\hline \hline

\textbf{Library}                & AES-CBC    & AES-GCM   
\\ \hline \hline
\textbf{OpenABE~\cite{OpenABE}} & $\square$         & $\blacksquare$ \\  \hline
\textbf{Charm~\cite{Charm}}     & $\blacksquare$    & $\square$      \\ \hline
\textbf{Rabe~\cite{Rabe}}       & $\square$         & $\blacksquare$   \\ \hline 
\textbf{GoFE~\cite{GoFE}}       & $\blacksquare$    & $\square$   \\ 
\hline \hline

\end{tabular}
}
\end{center}
\end{table}

\added{AES-CBC and AES-GCM are secure symmetric ciphers, but AES-GCM provides authenticated encryption. Using AES-GCM provides confidentiality and guarantees the integrity of the \ac{CT}. This protection does not stop attackers from violating the integrity of the \ac{CT} but allows the receiver to detect tampering. Distributed environments benefit from data integrity protection and should favor the use of AES-GCM.}
\added{As can be seen from the libraries presented in Table \ref{tab:3}, the only ones implementing AES-GCM are \textit{OpenABE} and \textit{Rabe}. It should be noted that \textit{Rabe} has implemented AES-GCM in \textit{rabe-0.3.1}}\footnote{https://docs.rs/rabe/0.3.1/rabe/index.html}. \added{Previous versions used the low-level AES block cipher function. Those versions should only be applied in academic and research environments and avoided for production.}

%% file: Docs/5.Experimental_evaluation_definition.tex
\section{Quantitative Evaluation Definition} \label{Sec:5}

\added{Following the qualitative evaluation, this section experimentally analyzes the four libraries identified as the most relevant from a developer's point of view. As mentioned above, one of the challenges developers face when implementing \ac{ABE} schemes is the lack of knowledge about the performance capabilities of the libraries. Therefore, this section studies and compares how libraries behave on devices with different computational capabilities. This behavior is quantified by measuring the time that each scheme of each library requires to perform the basic operations in \ac{ABE}: encryption, decryption, key generation, and, in \ac{dCP-ABE}, authority setup.}

\subsection{Experiment Definition} \label{Sec:5.1}

\added{Our experiment relies on the different types of ABE schemes. The classification of \ac{CP-ABE}, \ac{KP-ABE}, and \ac{dCP-ABE} is shown in Table \ref{tab:2}. \ac{dCP-ABE} schemes have a similar behavior to the conventional \ac{CP-ABE} schemes, but the key generation is distributed among several key-generating authorities.}

\added{The performance comparison of different \ac{ABE} schemes illustrates how practical aspects of the implementation, like the mathematical library or the implementation language, can affect \ac{ABE} schemes. Although this performance variation is predictable, it should be quantified. Therefore, the purpose is to provide developers with practical information about the implementation of different schemes and libraries to help them in their selection. Developers should consider that in the case of BDABE, the experiments only measure the time required for the scheme to perform cryptographic operations. All the time related to Blockchain operations is not considered here since it is highly variable and out of the scope of this paper.}

\deleted{Cryptographic libraries are affected by the scheme's complexity, their implemented language, and the mathematical library. These differences may result in execution times different from those expected, relevant for developers.}

\subsection{Testbed Setup} \label{Sec:5.2}

\replaced{The testbed consists of two \ac{RPI}, i.e., a \ac{RPI0} and a \ac{RPI4}. The \ac{RPI0} runs Raspbian Stretch, has 512MB of RAM, 1GHz, a single-core ARMv6, and Wi-Fi. We consider it a good representation of an \ac{IoT} device. The \ac{RPI4} has an ARMv8 processor, 8GB of RAM, and runs 32-bit Ubuntu Server TLS. This second device is considered representative of IoT devices with high processing capabilities.}{The testbed is formed by a \ac{RPI4} with an ARMv8 processor, 8GB of RAM, and running 32-bit Ubuntu Server TLS.} This setup provides a representative insight into the execution time of the libraries on \ac{IoT} devices, as well as libraries' multi-architecture capabilities. \deleted{A detailed analysis is provided in Table \ref{tab:2}. It shows that \textit{CiFEr} has been discarded because it requires a 64-bit architecture instead of the 32-bit architecture of the \ac{RPI}s. Therefore, the experiments are limited to Charm, \textit{GoFE},\textit{Rabe}, and \textit{OpenABE}.}

\replaced{As explained in the discussion of AES schemes in Section \ref{Sec:4.2}, \ac{ABE} schemes are paired with symmetric ciphers. Therefore, in this section's experiments, AES encrypts a 43-byte \ac{PT}, and \ac{ABE} schemes encrypt the 256-bit AES key. The reason behind the small-sized \mbox{\ac{PT}} is that this paper evaluates the computational efficiency of different \ac{ABE} implementations, so a small \ac{PT} has been chosen to make the computational burden of \ac{ABE} more significant than that of AES.}{CP-ABE is used in hybrid mode for the tests, the standard approach to deploy ABE schemes. AES encrypts a 43-byte PT, and CP-ABE encrypts the 256-bit AES key. A small PT has been chosen, so the computational burden of CP-ABE is more significant than that of AES. This provides a representative view of the efficiency of each library.} This provides a representative view of the efficiency of each library. \added{For the mentioned time measurements, experiments have been carried out by benchmarking, which depends on the library and its implementation language:}

\begin{itemize}
    \item \added{OpenABE: The library itself offers its proprietary benchmark.}
    \item \added{Rabe: Criterion for Rust.}
    \item \added{Charm: Timeit for Python.}
    \item \added{GoFE: Golang benchmarking.}
\end{itemize}

\added{It is worth mentioning that\textit{Rabe} runs on Rust nightly, so the native Rust benchmark can be used. However, this is an unstable feature that always returns the time value in nanoseconds. In complex schemes, the execution time is longer than the maximum value that can be returned by the u32 type variable used by the native benchmark. Therefore, an overflow is obtained which does not capture the actual duration of the function. Criterion, which is more stable and accurate, has been used instead.}

%% file: Docs/6.Results.tex
\section{Results} \label{Sec:6}

\added{To properly discuss the results, we present Table \ref{tab:4}, which summarizes the properties of each scheme. Most ABE schemes have a linear time growth in encryption and decryption. However, some offer unique features such as constant times for certain operations.}

\begin{table}[!htb]
\renewcommand{\arraystretch}{1.1}
\caption{Scheme features} \label{tab:4}
\begin{center}
\resizebox{0.94\columnwidth}{!}{
\begin{tabular}{lllll}
\hline \hline
Scheme
& \begin{tabular}[c]{@{}l@{}} Type \\ \fbox{\tiny{1}} \fbox{\tiny{2}} \fbox{\tiny{3}} \end{tabular}
& \begin{tabular}[c]{@{}l@{}} KeyGen \\ \fbox{\tiny{1}} \fbox{\tiny{2}} \end{tabular}
& \begin{tabular}[c]{@{}l@{}} Enc. \\ \fbox{\tiny{1}} \fbox{\tiny{2}} \end{tabular}
& \begin{tabular}[c]{@{}l@{}} Dec. \\ \fbox{\tiny{1}} \fbox{\tiny{2}} \end{tabular} \\

\hline \hline

& \begin{tabular}[c]{@{}l@{}} 1. KP-ABE \\ 2. CP-ABE \\ 3. dCP-ABE  \end{tabular} 
& \begin{tabular}[c]{@{}l@{}} 1. Linear \\ 2. Constant \end{tabular} 
& \begin{tabular}[c]{@{}l@{}} 1. Linear \\ 2. Constant  \end{tabular} 
& \begin{tabular}[c]{@{}l@{}} 1. Linear \\ 2. Constant  \end{tabular} \\

\hline \hline

GPSW06~\cite{10.1145/1180405.1180418}
& $\blacksquare$\hspace{6.5pt}$\square$\hspace{6.5pt}$\square$
& $\blacksquare$\hspace{6.5pt}$\square$
& $\blacksquare$\hspace{6.5pt}$\square$
& $\blacksquare$\hspace{6.5pt}$\square$
\\ \hline 
 
LSW10~\cite{5504791}
& $\blacksquare$\hspace{6.5pt}$\square$\hspace{6.5pt}$\square$
& $\blacksquare$\hspace{6.5pt}$\square$
& $\blacksquare$\hspace{6.5pt}$\square$
& $\blacksquare$\hspace{6.5pt}$\square$
\\ \hline

YCT14~\cite{YAO2015104}
& $\blacksquare$\hspace{6.5pt}$\square$\hspace{6.5pt}$\square$
& $\blacksquare$\hspace{6.5pt}$\square$
& $\blacksquare$\hspace{6.5pt}$\square$
& $\blacksquare$\hspace{6.5pt}$\square$
\\ \hline

FAME~\cite{10.1145/3133956.3134014}
& $\blacksquare$\hspace{6.5pt}$\square$\hspace{6.5pt}$\square$
& $\blacksquare$\hspace{6.5pt}$\square$
& $\blacksquare$\hspace{6.5pt}$\square$
& $\square$\hspace{6.5pt}$\blacksquare$
\\ \hline \hline

BSW07~\cite{4223236}
& $\square$\hspace{6.5pt}$\blacksquare$\hspace{6.5pt}$\square$
& $\blacksquare$\hspace{6.5pt}$\square$
& $\blacksquare$\hspace{6.5pt}$\square$
& $\blacksquare$\hspace{6.5pt}$\square$
\\ \hline

W11~\cite{waters2011ciphertext}
& $\square$\hspace{6.5pt}$\blacksquare$\hspace{6.5pt}$\square$
& $\blacksquare$\hspace{6.5pt}$\square$
& $\blacksquare$\hspace{6.5pt}$\square$
& $\blacksquare$\hspace{6.5pt}$\square$
\\ \hline

YAHK14~\cite{10.1007/978-3-642-54631-0_16}
& $\square$\hspace{6.5pt}$\blacksquare$\hspace{6.5pt}$\square$
& $\blacksquare$\hspace{6.5pt}$\square$
& $\blacksquare$\hspace{6.5pt}$\square$
& $\blacksquare$\hspace{6.5pt}$\square$
\\ \hline

FAME~\cite{10.1145/3133956.3134014}
& $\square$\hspace{6.5pt}$\blacksquare$\hspace{6.5pt}$\square$
& $\blacksquare$\hspace{6.5pt}$\square$
& $\blacksquare$\hspace{6.5pt}$\square$
& $\square$\hspace{6.5pt}$\blacksquare$
\\ \hline \hline

MKE08~\cite{10.1007/978-3-642-00730-9_2}
& $\square$\hspace{6.5pt}$\square$\hspace{6.5pt}$\blacksquare$
& $\blacksquare$\hspace{6.5pt}$\square$
& $\square$\hspace{6.5pt}$\blacksquare$
& $\square$\hspace{6.5pt}$\blacksquare$
\\ \hline

LW11~\cite{lewko2011decentralizin}
& $\square$\hspace{6.5pt}$\square$\hspace{6.5pt}$\blacksquare$
& $\blacksquare$\hspace{6.5pt}$\square$
& $\blacksquare$\hspace{6.5pt}$\square$
& $\blacksquare$\hspace{6.5pt}$\square$
\\ \hline

RW15~\cite{10.1007/978-3-662-47854-7_19}
& $\square$\hspace{6.5pt}$\square$\hspace{6.5pt}$\blacksquare$
& $\blacksquare$\hspace{6.5pt}$\square$
& $\blacksquare$\hspace{6.5pt}$\square$
& $\blacksquare$\hspace{6.5pt}$\square$
\\ \hline

BDABE~\cite{secrypt18}
& $\square$\hspace{6.5pt}$\square$\hspace{6.5pt}$\blacksquare$
& $\blacksquare$\hspace{6.5pt}$\square$
& $\square$\hspace{6.5pt}$\blacksquare$
& $\square$\hspace{6.5pt}$\blacksquare$ 

\\ \hline \hline

\end{tabular}
}
\end{center}
\end{table}


\subsection{CP-ABE} \label{Sec:6.1}

\added{For this experiment, we analyze the time evolution of the different operations (i.e., keygen, encryption, and decryption) for a different number of attributes. This way, we quantify the variance of the time requirements according to the attributes contained in the \ac{AP} or \ac{SK}.}

\begin{figure}[!h] 
\centering
\begin{subfigure}{1\columnwidth}
    \includegraphics[width=1\textwidth]{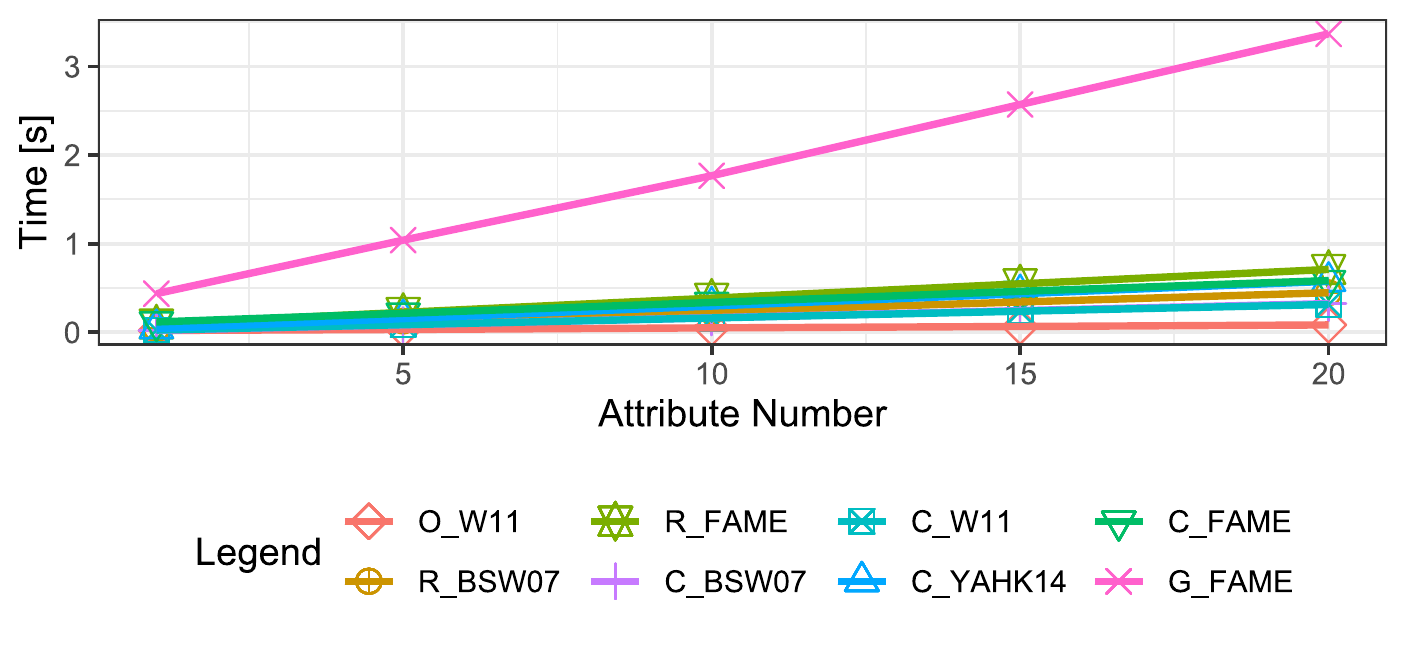}
    \caption{Results in a \ac{RPI4}}
    \label{fig:CPABE_KeyGen_GOFE-RPI4}
\end{subfigure}
\hfill
\begin{subfigure}{1\columnwidth}
    \includegraphics[width=1\textwidth]{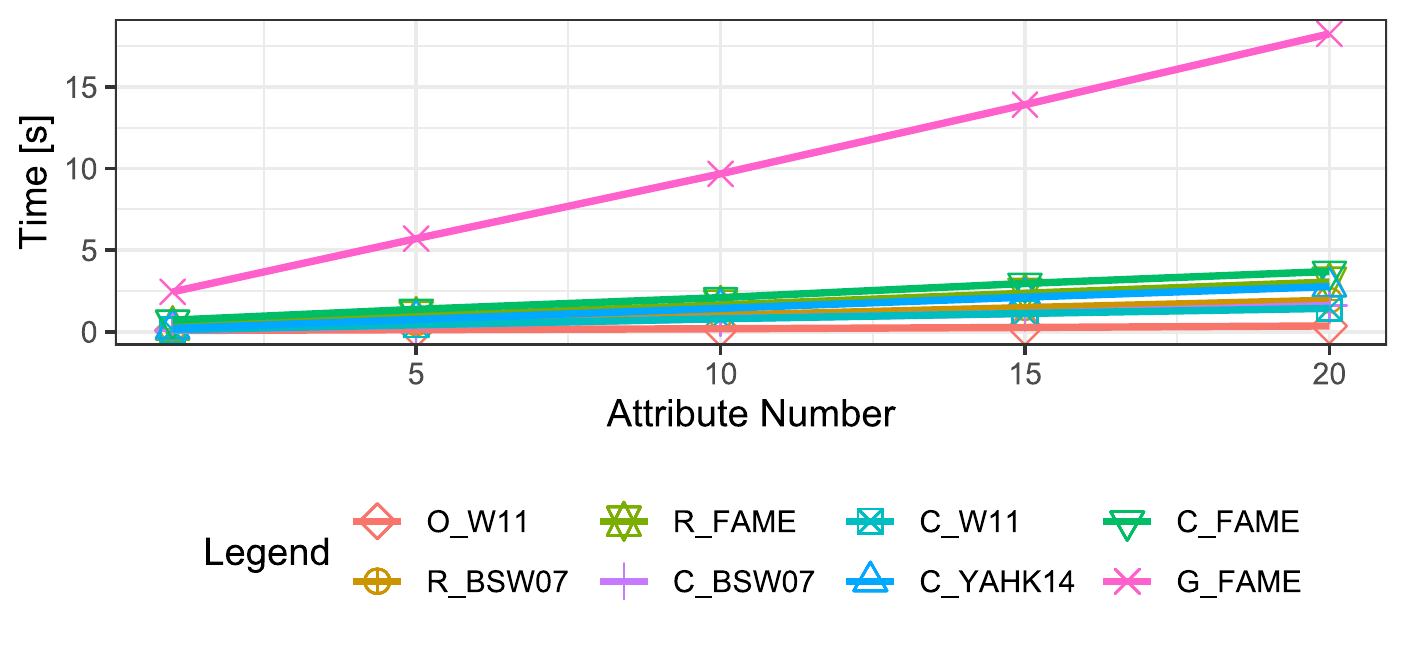}
    \caption{Results in a \ac{RPI0}}
    \label{fig:CPABE_KeyGen_GOFE-RPI0}
\end{subfigure}

\caption{CP-ABE Key Generation times in seconds.}
\label{fig:CPABE_KeyGen_GOFE}
\end{figure}

\replaced{Figure \ref{fig:CPABE_KeyGen_GOFE} shows the time in seconds required to generate secret keys for a different number of attributes.}{Figure \ref{fig:CPABE_KeyGen_GOFE} shows the time in seconds it takes to generate secret keys according to a fixed attribute number.} \added{The required time grows linear with the number of attributes used for the \ac{SK} generation.} \replaced{GoFE-FAME (G\_FAME in the Figure) is the slowest implementation for key generation. It takes 3.36s in the \ac{RPI4} and 18s in the \ac{RPI0}. Meanwhile, Rabe-FAME (R\_FAME) takes 0.71s for the worst case in the \ac{RPI4} and 3s in the \ac{RPI0}.}{\textit{GoFE} is the slowest implementation for key generation, taking 3.35s, while\textit{Rabe} takes 0.88s for the worst case. This is 280\% more time than\textit{Rabe} for the same operation.}

\begin{figure}[!h] 
\centering
\begin{subfigure}{0.95\columnwidth}
    \includegraphics[width=1\textwidth]{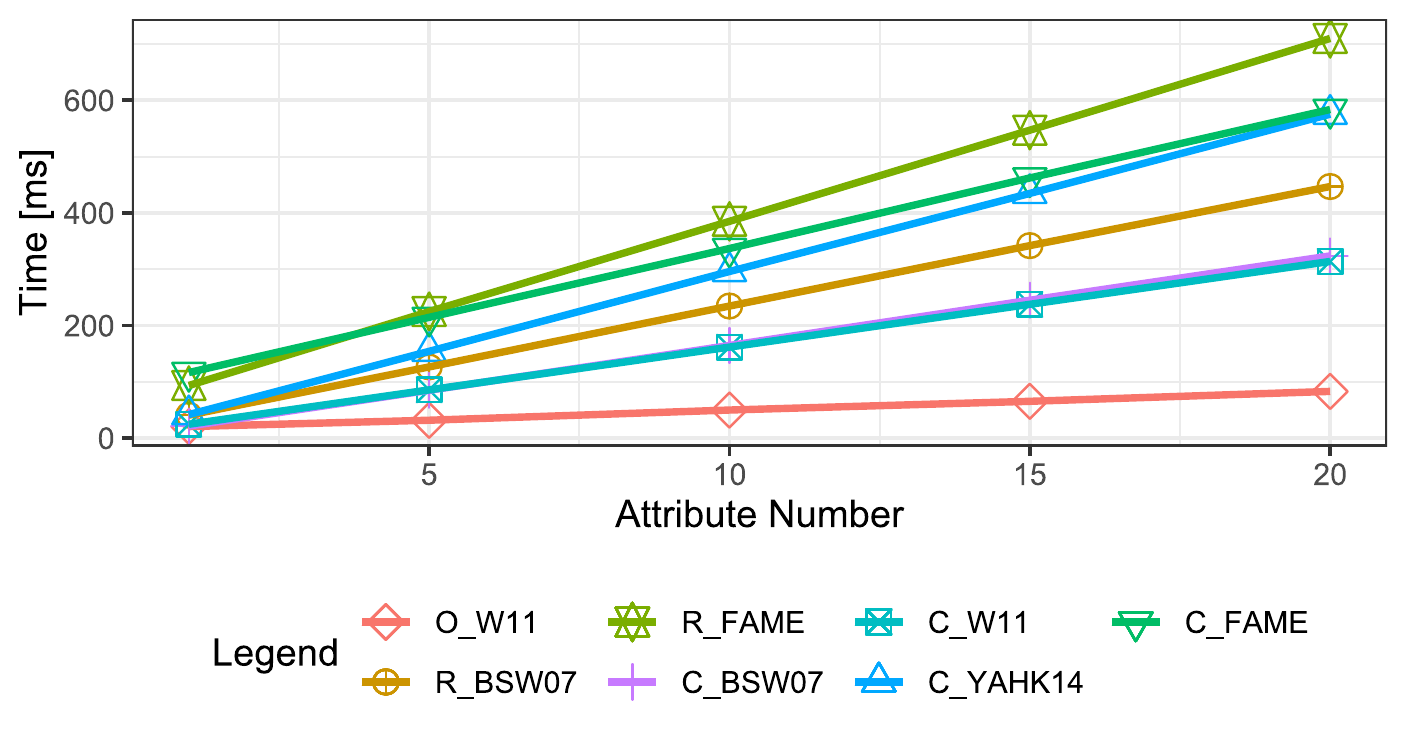}
    \caption{Results in a \ac{RPI4}}
    \label{fig:CPABE_KeyGen_sinGOFE-RPI4}
\end{subfigure}
\hfill
\begin{subfigure}{0.95\columnwidth}
    \includegraphics[width=1\textwidth]{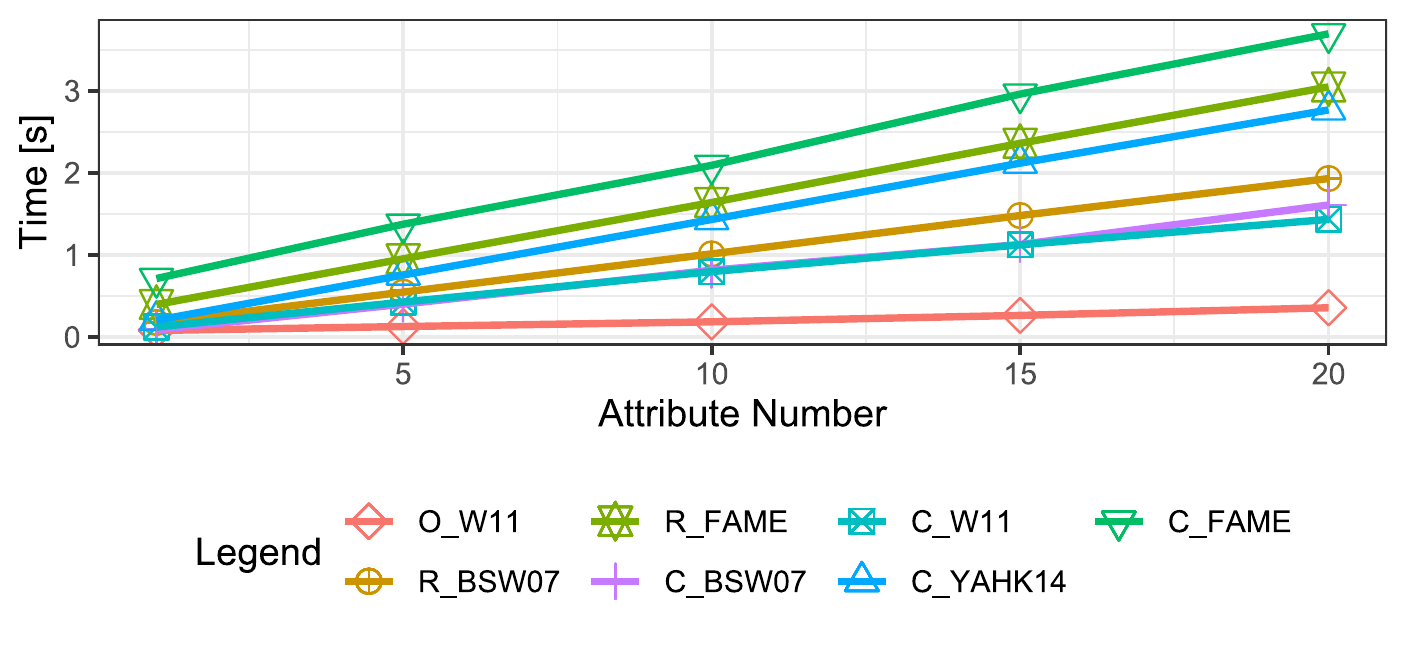}
    \caption{Results in a \ac{RPI0}}
    \label{fig:CPABE_KeyGen_sinGOFE-RPI0}
\end{subfigure}
\caption{CP-ABE Key Generation times in seconds without G\_FAME.}
\label{fig:CPABE_KeyGen_sinGOFE}
\end{figure}

The difference \added{in times shown in Figure \ref{fig:CPABE_KeyGen_GOFE}} is related to \textit{GoFE} using crypt-go, Golang's cryptographic libraries, which are less efficient than consolidated libraries like OpenSSL. There are forks of Go that add additional security features\footnote{https://github.com/cloudflare/go}, but experiments for this paper have been carried out with the official Go distribution. Meanwhile, \textit{Charm} and \textit{OpenABE} use well-established and consolidated math libraries like PBC, Mircl, or Relic. These libraries have had numerous releases and patches that have added functionality and increased efficiency. \deleted{We can see that among the three other libraries,\textit{Rabe} is the slowest one taking 0.88s to the 0.083s of \textit{OpenABE}. \textit{Charm} lies in-between, taking 0.58s.} \added{Since programming languages are designed for different purposes, their optimization level impacts performance. However, the impact of mathematical dependencies is also noticeable: Figure \ref{fig:CPABE_KeyGen_GOFE} shows that a library written in an interpreted language (i.e., Python) is faster than one written in a compiled language (e.g., Go or Rust). The growth rate of \textit{GoFE} makes it challenging to visualize the rest of the results. Therefore, we provide Figure \ref{fig:CPABE_KeyGen_sinGOFE} for a better view.}

\added{Figure \ref{fig:CPABE_KeyGen_sinGOFE} shows that the fastest scheme for key generation is OpenABE-Waters11 (O\_W11), which takes 83ms for twenty attributes in the \ac{RPI4} and 355ms in the \ac{RPI0}. The next best scheme is Charm-Waters11 (C\_W11), with a notable difference from O\_W11. In fact, O\_W11 is approximately 74\% faster than CW11 in both \ac{RPI4} and \ac{RPI0}.}

\added{Another interesting finding is that Charm-BSW07 (C\_BSW07) takes a similar time as C\_W11 in both devices. In the \ac{RPI4}, C\_BSW07 takes 327ms for the worst case, barely 13ms longer than C\_W11. In the case of \ac{RPI0}, C\_BSW07 takes 1.60s and C\_W11 1.43s. Thus, the difference between C\_BSW07 and C\_W11 is 4.375\% in \ac{RPI4} and 11.22\% in \ac{RPI0}.}

\added{Once we establish G\_FAME as the slowest scheme, it is interesting to see which one is the second slowest scheme. In \ac{RPI4}, the second slowest scheme is R\_FAME at 0.70988s. Meanwhile, in the \ac{RPI0}, the second slowest scheme is C\_FAME at 3.69s. This is because \ac{RPI0} favors compiled languages like Rust and it is more affected by interpreted languages like Python.}

\begin{figure}[!htb] 
\centering
\begin{subfigure}{1\columnwidth}
    \includegraphics[width=1\textwidth]{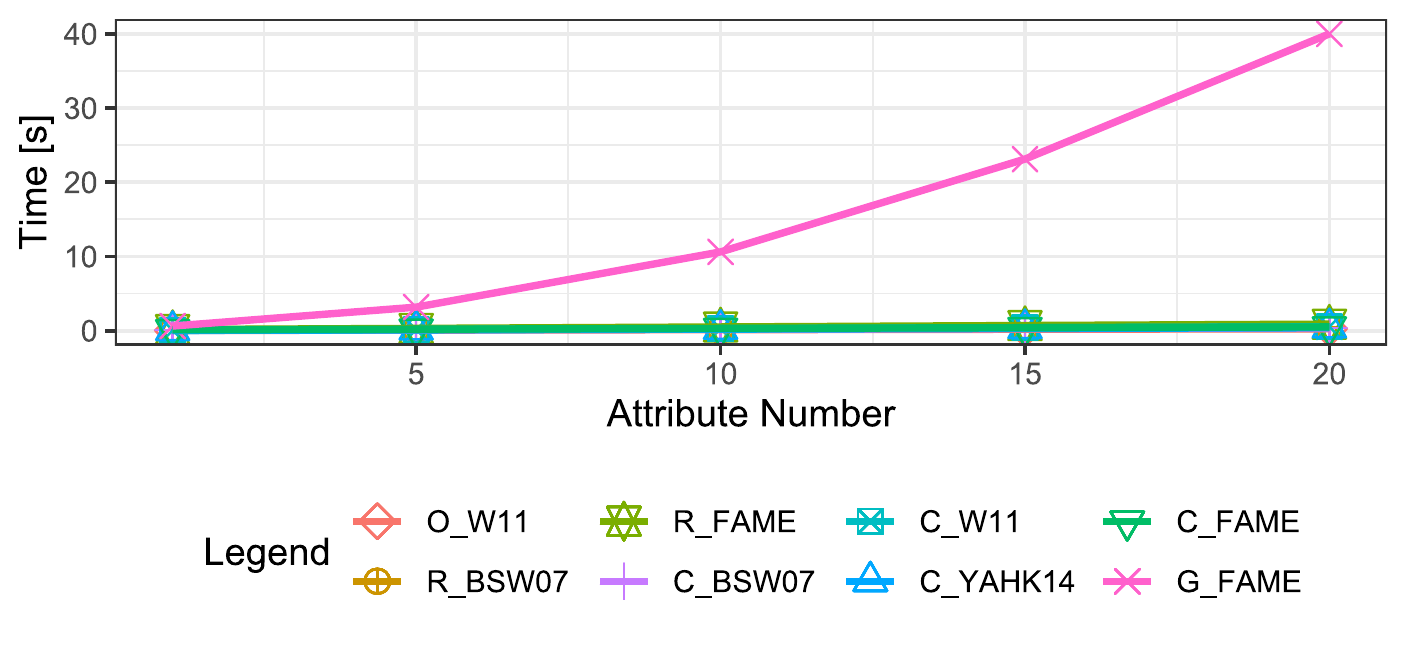}
    \caption{Results in a \ac{RPI4}}
    \label{fig:CPABE_Enc_conGOFE-RPI4}
\end{subfigure}
\hfill
\begin{subfigure}{1\columnwidth}
    \includegraphics[width=1\textwidth]{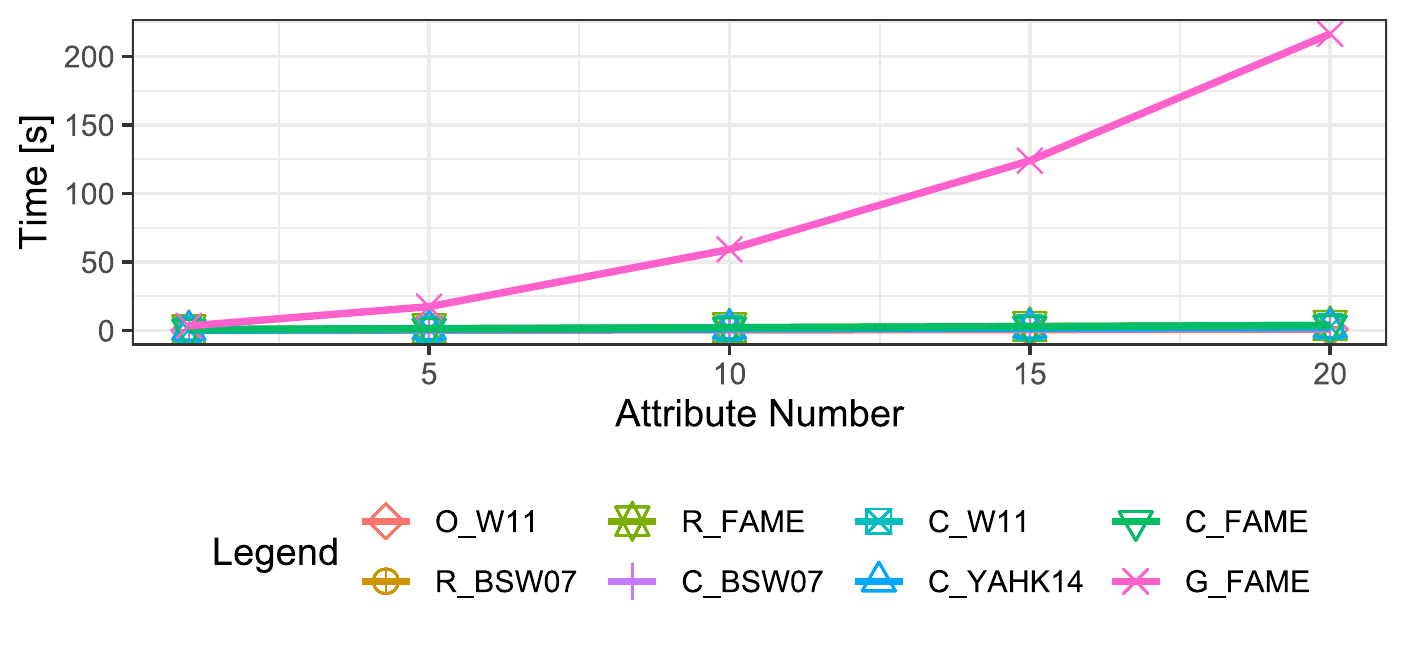}
    \caption{Results in a \ac{RPI0}}
    \label{fig:CPABE_Enc_conGOFE-RPI0}
\end{subfigure}
\caption{CP-ABE encryption times in seconds.}
\label{fig:CPABE_Enc_conGOFE}
\end{figure}

\begin{figure}[!h] 
\centering
\begin{subfigure}{0.97\columnwidth}
    \includegraphics[width=1\textwidth]{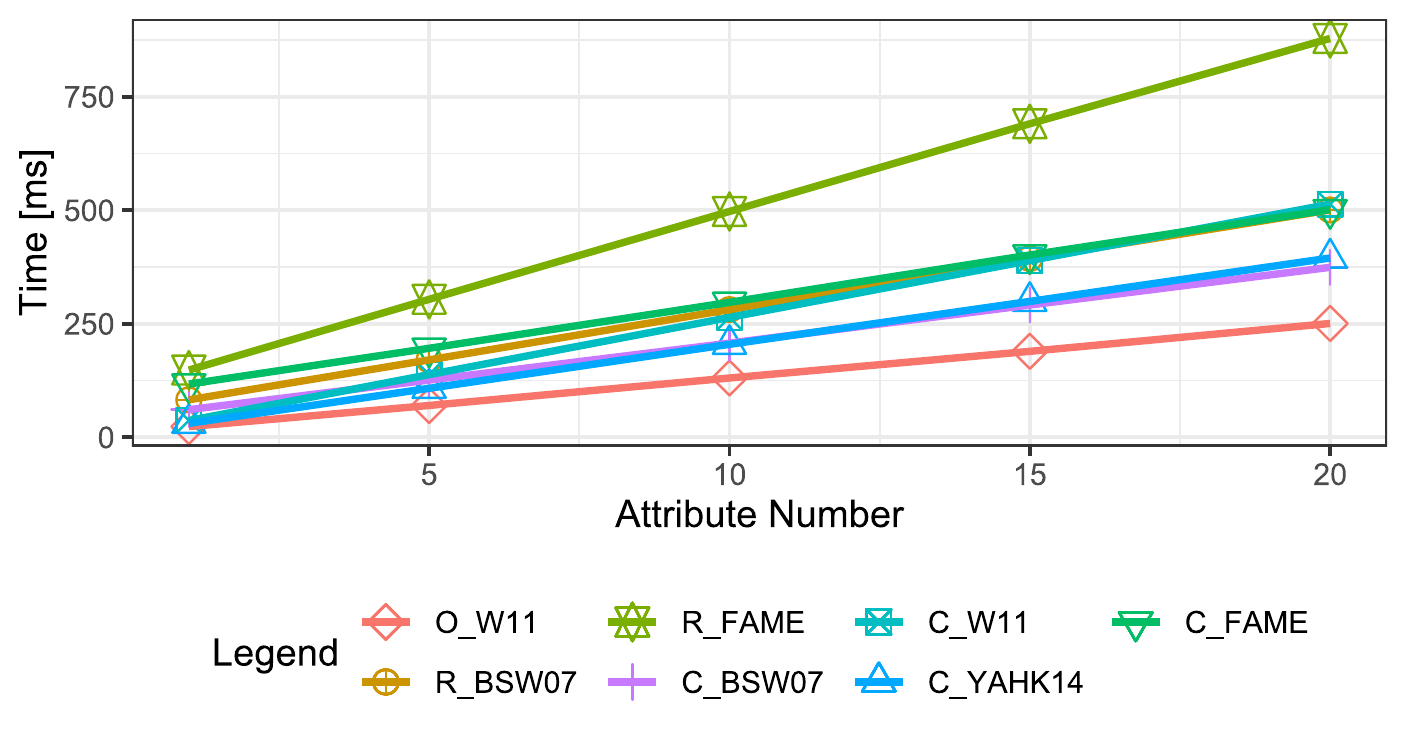}
    \caption{Results in a \ac{RPI4}}
    \label{fig:CPABE_Enc_sinGOFE-RPI4}
\end{subfigure}
\hfill
\begin{subfigure}{0.97\columnwidth}
    \includegraphics[width=1\textwidth]{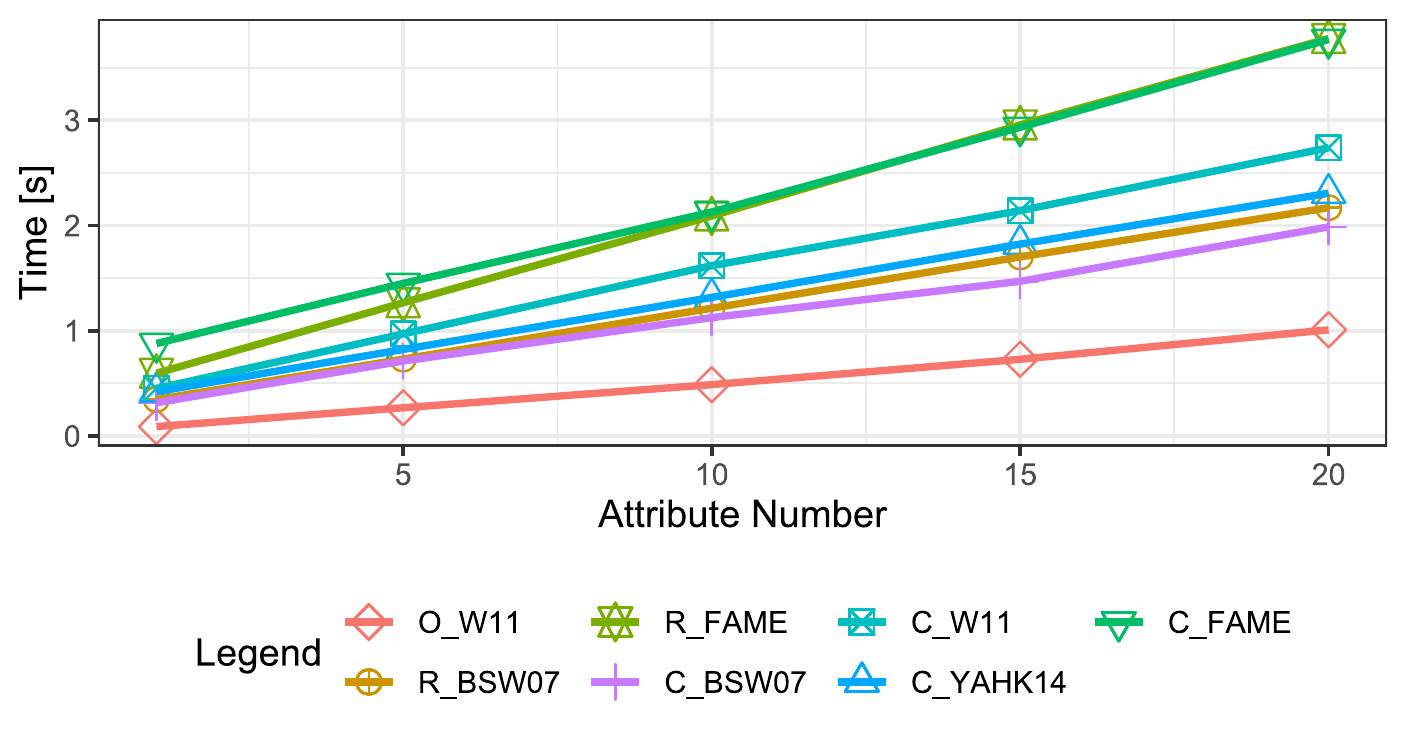}
    \caption{Results in a \ac{RPI0}}
    \label{fig:CPABE_Enc_sinGOFE-RPI0}
\end{subfigure}
\caption{CP-ABE encryption times in seconds, without G\_FAME.}
\label{fig:CPABE_Enc_sinGOFE}
\end{figure}

\added{Key generation is critical, and its running time should be measured. However, this operation is performed at a much lower frequency than encryption and decryption. To study the libraries' encryption times, we present Figure \ref{fig:CPABE_Enc_conGOFE}. It shows that G\_FAME grows exponentially, taking 40 seconds for the worst case in the \ac{RPI4} and 3 minutes for the \ac{RPI0}. Meanwhile, the rest of the library-scheme combinations take less than a second in the \ac{RPI4} and less than four seconds in \ac{RPI0}. Furthermore, FAME encryption time grows linearly with the complexity of the \ac{AP}, which G\_FAME fulfills in neither device. Because of the long time taken by \textit{GoFE} to encrypt information, we present Figure \ref{fig:CPABE_Enc_sinGOFE} to be able to visualize the rest of the results.} 

\deleted{In the case of encryption, Figure \ref{fig:CPABE_Enc_conGOFE-RPI4} exponentially over time, taking as long as 40s to encrypt the \ac{PT}, while Charm,\textit{Rabe}, and \textit{OpenABE} take less than a second. }

\added {We can see in Figure \ref{fig:CPABE_Enc_sinGOFE-RPI4} that R\_FAME is the slowest encryption scheme for every case in the \ac{RPI4}. At 15 attributes, it holds a notable difference from the next-slowest one, C\_FAME. For this case, R\_FAME takes 690.36ms and C\_FAME 400ms, making R\_FAME 72.5\% slower. However, this changes for more than 15 attributes, and C\_W11 becomes the4 second slowest scheme. At 20 attributes, C\_W11 takes 513.42ms and R\_FAME 878.26, making this last scheme 70\% slower. Developers should keep the dependence on the \ac{AP} complexity in mind when implementing CP-ABE schemes. The fastest scheme is O\_W11, which takes 250 ms for 20 attributes, making it 33\% faster than the second fastest scheme (C\_BSW07) for the same case.}

\added{Figure \ref{fig:CPABE_Enc_sinGOFE-RPI0} presents the results for the \ac{RPI0}. It shows that C\_FAME and R\_FAME have a slight difference of 0.26\% in the case of 20 attributes: 3.76s for C\_FAME and 3.77s for R\_FAME. Overall, the fastest scheme is O\_W11, which takes 1s for the case of 20 attributes. Due to being implemented in C++, it is highly efficient and more suitable for the \ac{RPI0} than the python version of the same scheme (i.e., C\_W11). Regarding the second fastest scheme, C\_BSW07 takes 1.98s for the worst case. This makes the difference between C\_BSW07 and O\_W11 noteworthy since it makes O\_W11 49\% faster.} 

\begin{figure}[!h] 
\centering
\begin{subfigure}{0.95\columnwidth}
    \includegraphics[width=1\textwidth]{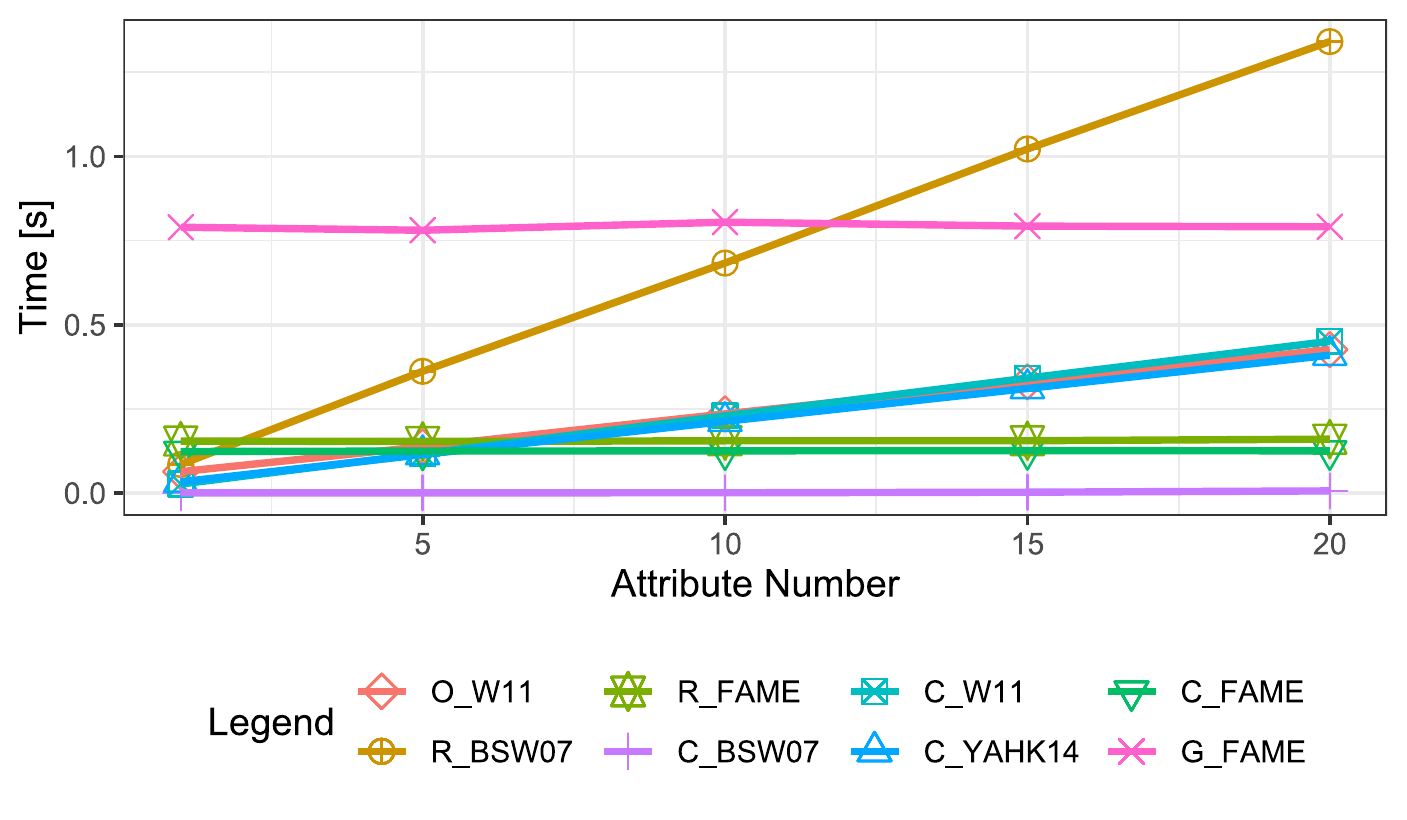}
    \caption{Results in a \ac{RPI4}}
    \label{fig:CPABE_DEC-RPI4}
\end{subfigure}
\hfill
\begin{subfigure}{0.95\columnwidth}
    \includegraphics[width=1\textwidth]{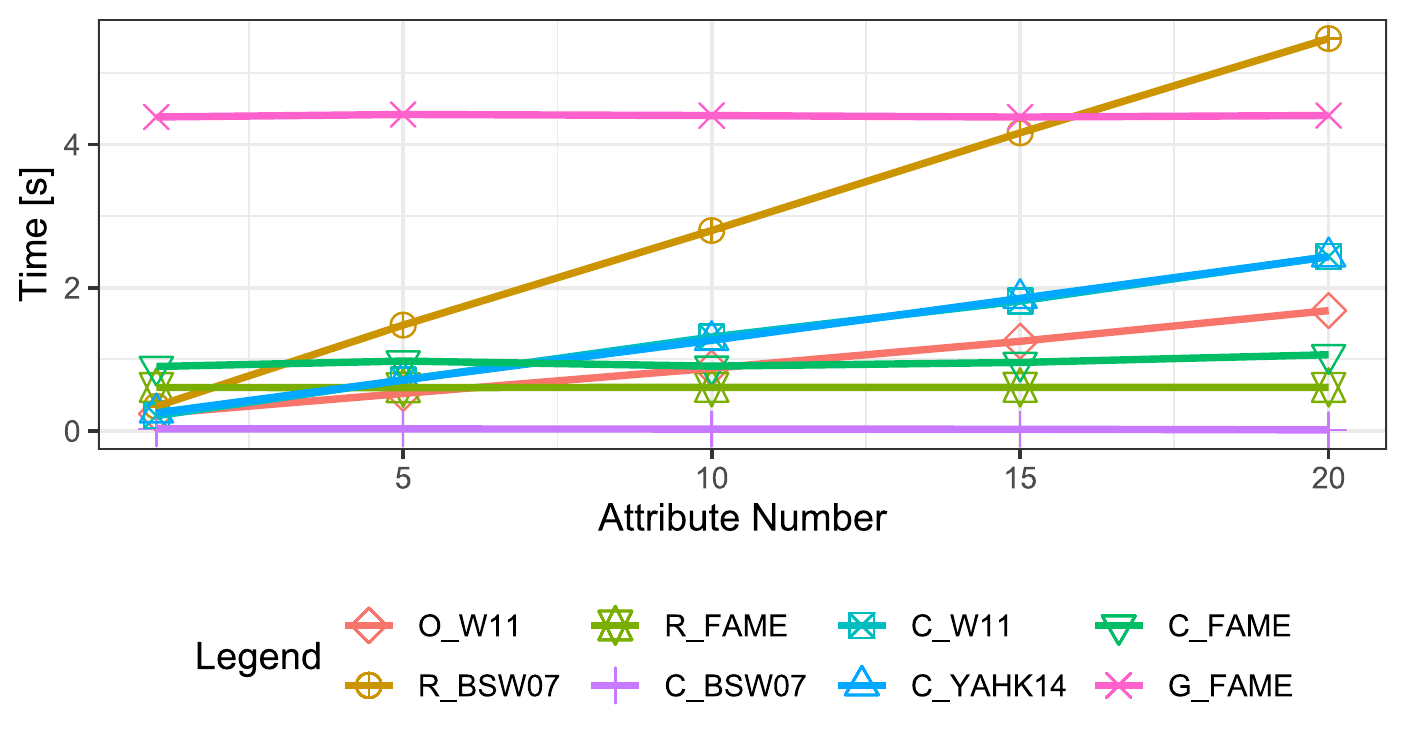}
    \caption{Results in a \ac{RPI0}}
    \label{fig:CPABE_DEC-RPI0}
\end{subfigure}
\caption{CP-ABE decryption times in seconds.}
\label{fig:CPABE_DEC}
\end{figure}

\added{Although encryption times grow linearly with the number of attributes contained in the \ac{CT}, the pattern does not always hold for decryption. As was introduced in Table \ref{tab:4} some schemes (e.g., FAME) have the distinctive feature of constant decryption time, independent of the number of attributes in the \ac{CT}. This is seen in Figure \ref{fig:CPABE_DEC}, where G\_FAME decryption takes 800ms for every case in a \ac{RPI4} and approximately 4s in a \ac{RPI0}. However, the Python implementation (C\_FAME) and Rust implementation (R\_FAME) take 160ms and 130ms, approximately 80\%-84\% less time than G\_FAME.}


\added{One of the differences between the RPI and RPI4 is that in \ac{RPI0}, C\_FAME and R\_FAME no longer have similar times. In fact, Rust's efficiency over Python is apparent in these results, with R\_FAME taking 607.69ms and C\_FAME taking approximately 950ms, 56\% more time for the same scheme.}

\added{Interestingly, G\_FAME is not always the slowest scheme. In the case of \ac{RPI4}, the slowest scheme for more than 12 attributes is R\_BSW07, taking up to 1.34s to decrypt a \ac{CT} of 20 attributes. In the case of \ac{RPI0}, the same happens for more than 16 attributes, with R\_BSW07 taking 4.47s for a \ac{CT} of 20 attributes. Finally, the fastest option in both devices is C\_BSW07.}

\deleted{A similar situation occurs with decryption, Figure \ref{fig:CPABE_DEC-RPI4}.} 



\deleted{To observe better which library is faster (Rabe, Charm, or \textit{OpenABE}), Figure \ref{fig:CPABE_ENCDEC-RPI4} depicts the same results on encryption and decryption but without \textit{GoFE} this time. \textit{OpenABE} is the fastest scheme during encryption (solid green line), at 250ms. Regarding\textit{Rabe} and Charm, we can see that \textit{Charm} takes 500ms for 20 attributes while\textit{Rabe} takes 700ms. However, since \textit{OpenABE} implements W11-LU instead of FAME, decryption times vary. FAME has constant decryption times, while W11-LU scheme decryption time scale linearly with the number of attributes in the policy. As a result, \textit{OpenABE} is the fastest scheme for five attributes or less. For policies more complex than that, FAME scheme is better. In this context, \textit{Charm} performs better than\textit{Rabe}, although the difference is almost negligible, with \textit{Charm} taking 130ms and\textit{Rabe} 150ms.}

\subsection{KP-ABE} \label{Sec:6.2}

\begin{figure}[!ht] 
\centering
\begin{subfigure}{1\columnwidth}
    \includegraphics[width=1\textwidth]{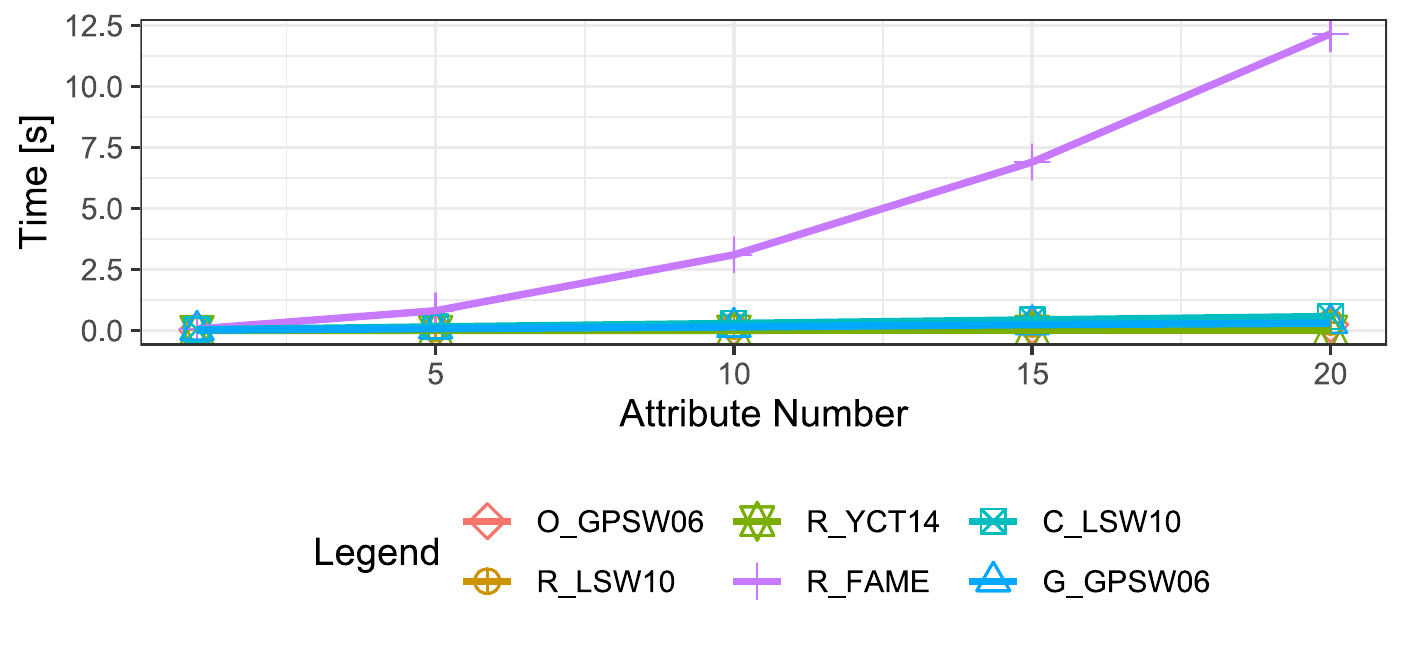}
    \caption{Results in a \ac{RPI4}}
    \label{fig:KPABE_KeyGen_conFAME-RPI4}
\end{subfigure}
\hfill
\begin{subfigure}{1\columnwidth}
    \includegraphics[width=1\textwidth]{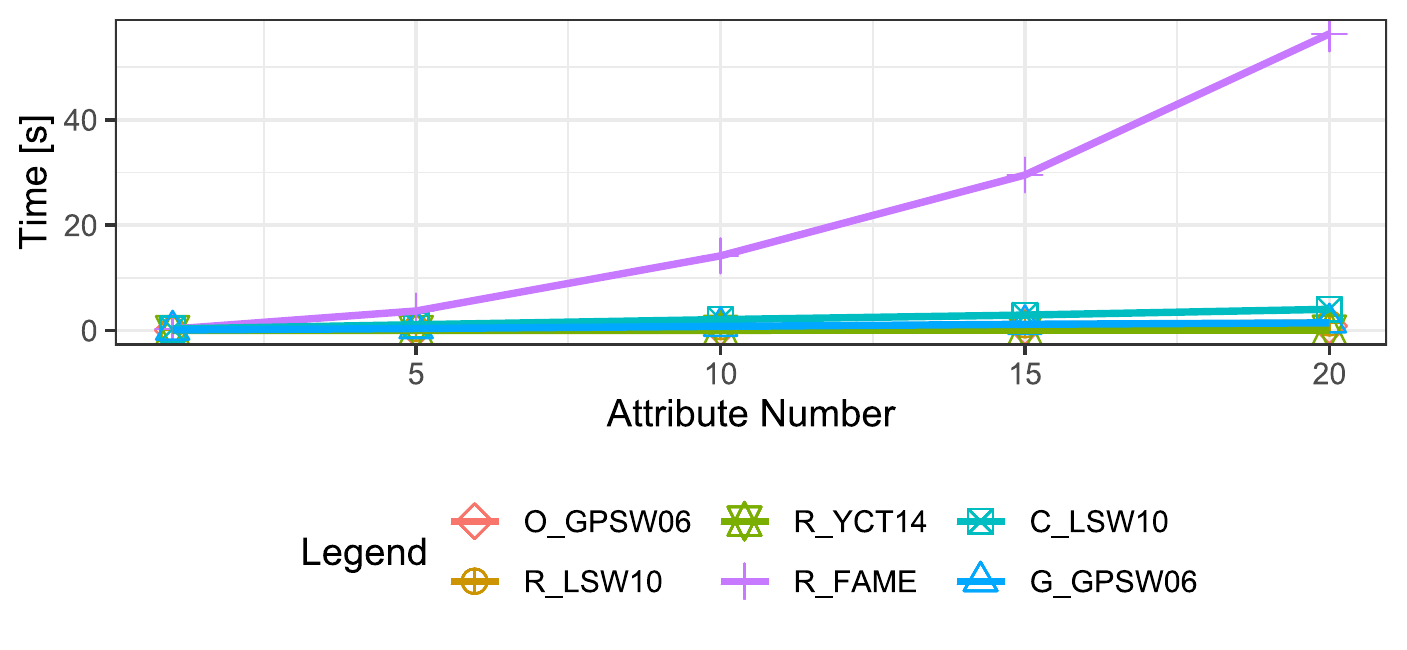}
    \caption{Results in a \ac{RPI0}}
    \label{fig:KPABE_KeyGen_conFAME-RPI0}
\end{subfigure}
\caption{KP-ABE Key Generation times in seconds.}
\label{fig:KPABE_KeyGen_conFAME}
\end{figure}

\added{This section conducts an analysis of libraries that implement KP-ABE schemes, analogous to that conducted for CP-ABE in Section \ref{Sec:6.1}. To this end, we present Figure \ref{fig:KPABE_KeyGen_conFAME}, which shows the time required to create users' secret keys. KP-ABE associates user keys with APs. Thus, this experiment analyzes the time required to generate the keys as a function of the number of attributes contained in the policy.}

\begin{figure}[!h] 
\centering
\begin{subfigure}{1\columnwidth}
    \includegraphics[width=1\textwidth]{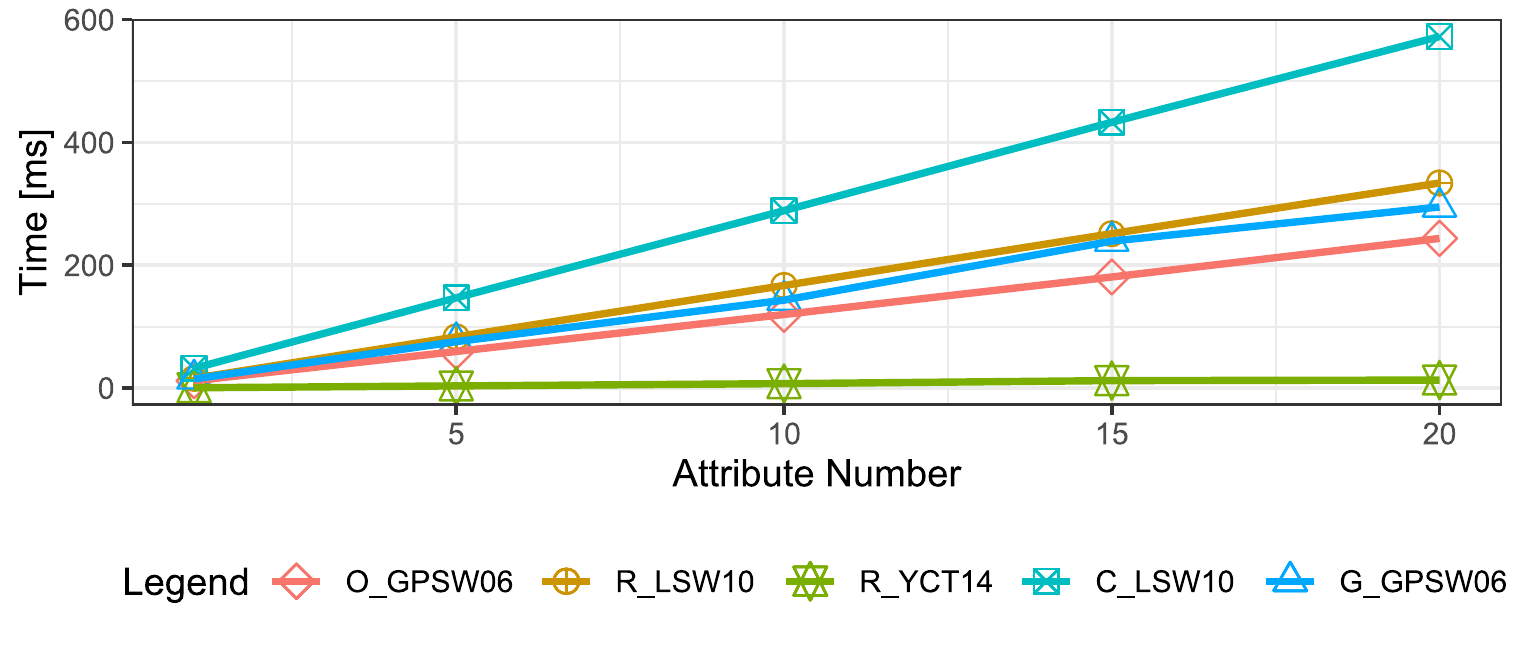}
    \caption{Results in a \ac{RPI4}}
    \label{fig:KPABE_KeyGen_sinFAME-RPI4}
\end{subfigure}
\hfill
\begin{subfigure}{1\columnwidth}
    \includegraphics[width=1\textwidth]{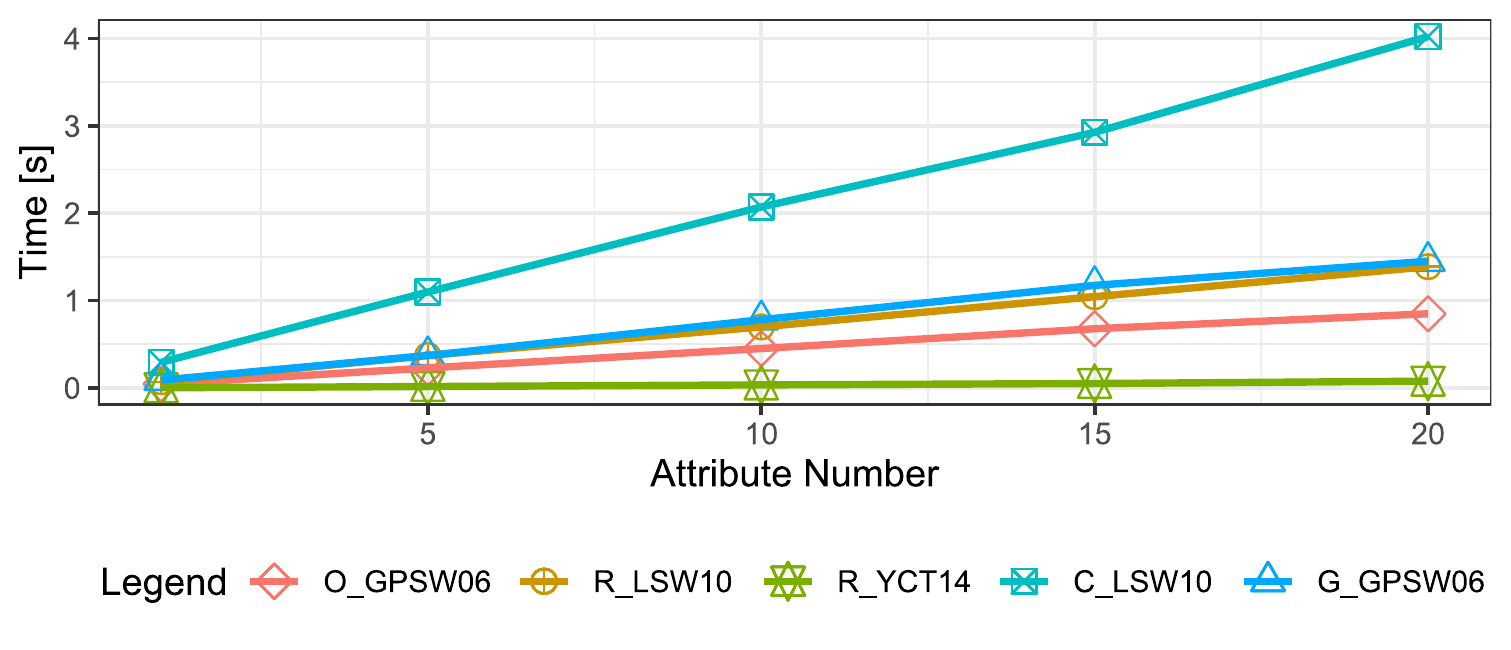}
    \caption{Results in a \ac{RPI0}}
    \label{fig:KPABE_KeyGen_sinFAME-RPI0}
\end{subfigure}
\caption{KP-ABE Key Generation times in seconds, without R\_FAME.}
\label{fig:KPABE_KeyGen_sinFAME}
\end{figure}

\added{Figure \ref{fig:KPABE_KeyGen_conFAME} shows that the FAME implementation of \textit{Rabe}, named (R\_FAME) in the figure, has an exponential time evolution for user key generation. In fact, for the case of 20 attributes, the time required goes up to 12s for \ac{RPI4} and up to 56s for \ac{RPI0}. The difference with the second slowest scheme is big enough that we provide Figure \ref{fig:KPABE_KeyGen_sinFAME} to visualize it.}

\added{Figure \ref{fig:KPABE_KeyGen_sinFAME} shows that R\_YCT14 is the fastest scheme-library combination. It takes 12ms to generate a key with 20 attributes for a \ac{RPI4} and 76.4ms for a \ac{RPI0}. In RPI4, R\_YCT14 shows a significant difference with the second fastest scheme, O\_GPSW07. For the same case of 20 attributes, O\_GPSW07 requires 243s, a difference of 181\%. Meanwhile, in the RPI0, O\_GPSW07 takes 848ms, a difference of 166\% from R\_YCT14.}

\added{Regarding the second slowest scheme, C\_LSW10, it is interesting to see that in the \ac{RPI4}, it requires 572ms. This implies 71\% more time than the same scheme in Rust (R\_LSW10), 334ms. However, as Figure \ref{fig:KPABE_KeyGen_sinFAME-RPI0} presents, in \ac{RPI0}, the difference between both schemes goes up to 185\%: 4s for the worst case in C\_LSW10 and 1.4s in the R\_LSW10. The experiment, thus, clearly shows how devices with reduced computing capabilities favor libraries written in Rust and C++ over those implemented in Python.}

\begin{figure}[!h] 
\centering
\begin{subfigure}{1\columnwidth}
    \includegraphics[width=1\textwidth]{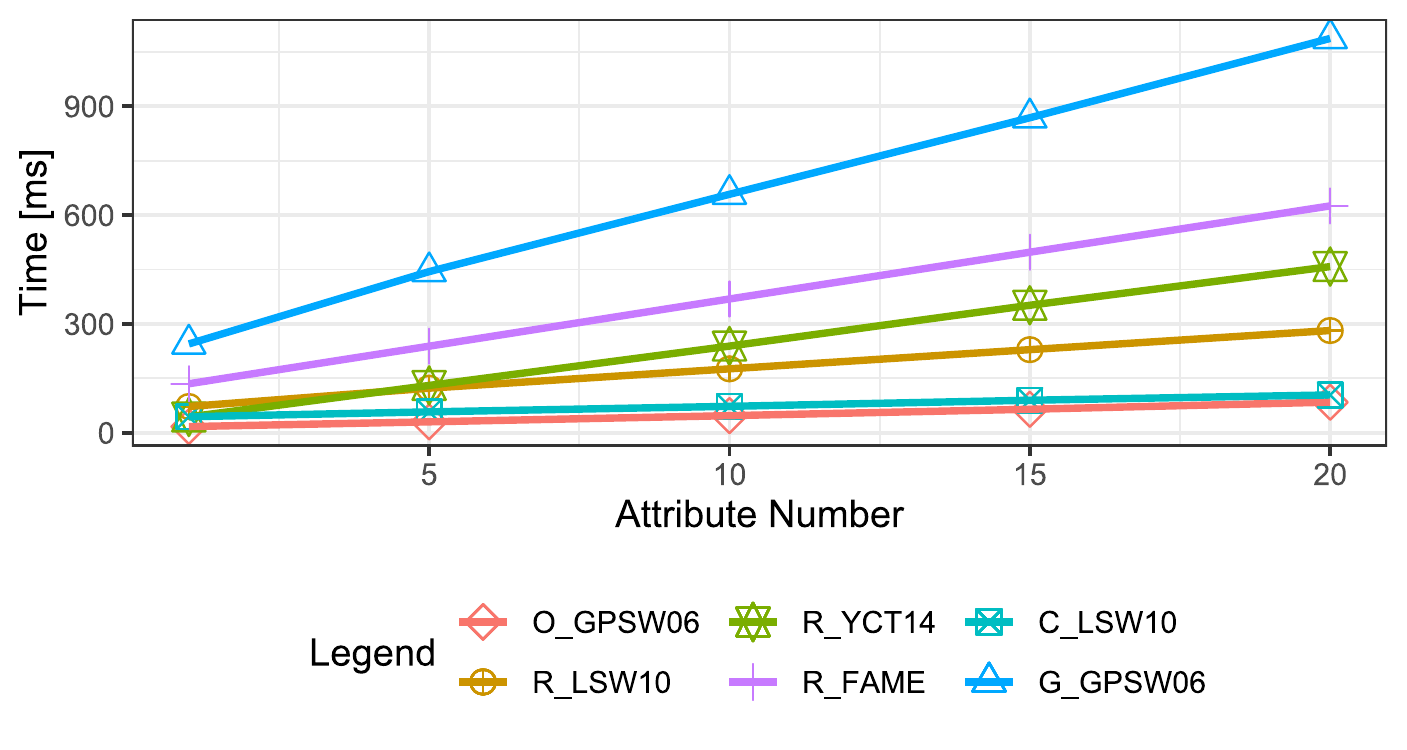}
    \caption{Results in a \ac{RPI4}}
    \label{fig:KPABE_Enc-RPI4}
\end{subfigure}
\hfill
\begin{subfigure}{1\columnwidth}
    \includegraphics[width=1\textwidth]{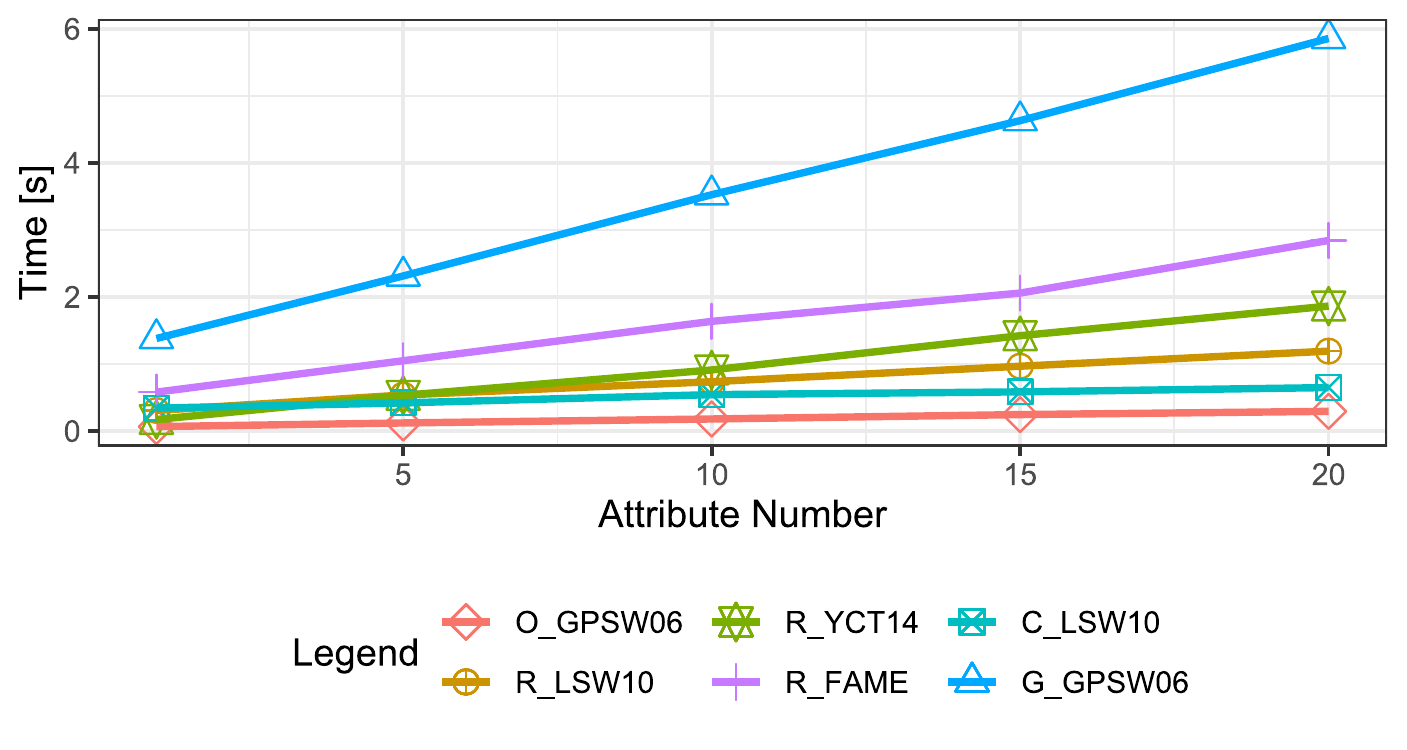}
    \caption{Results in a \ac{RPI0}}
    \label{fig:KPABE_Enc-RPI0}
\end{subfigure}
\caption{KP-ABE encryption times.}
\label{fig:KPABE_Enc}
\end{figure}

\added{When it comes to encryption, the results can be seen in Figure \ref{fig:KPABE_Enc}. It is clear from the figure that the fastest scheme for encryption is O\_GPSW06, taking 84ms for 20 attributes in a \ac{RPI4} and 292ms in a \ac{RPI0}. Despite its slow key generation times, C\_LSW10 is a close second for the most efficient encryption scheme. In a \ac{RPI4} it takes 104ms for the worst case: barely 23\% more than O\_GPSW06. In a \ac{RPI0}, however, C\_LSW10 requires 646ms for the worst case: 121\% more than O\_GPSW06. In contrast, the slowest scheme in both devices is G\_GPSW06.}

\begin{figure}[!h] 
\centering
\begin{subfigure}{1\columnwidth}
    \includegraphics[width=1\textwidth]{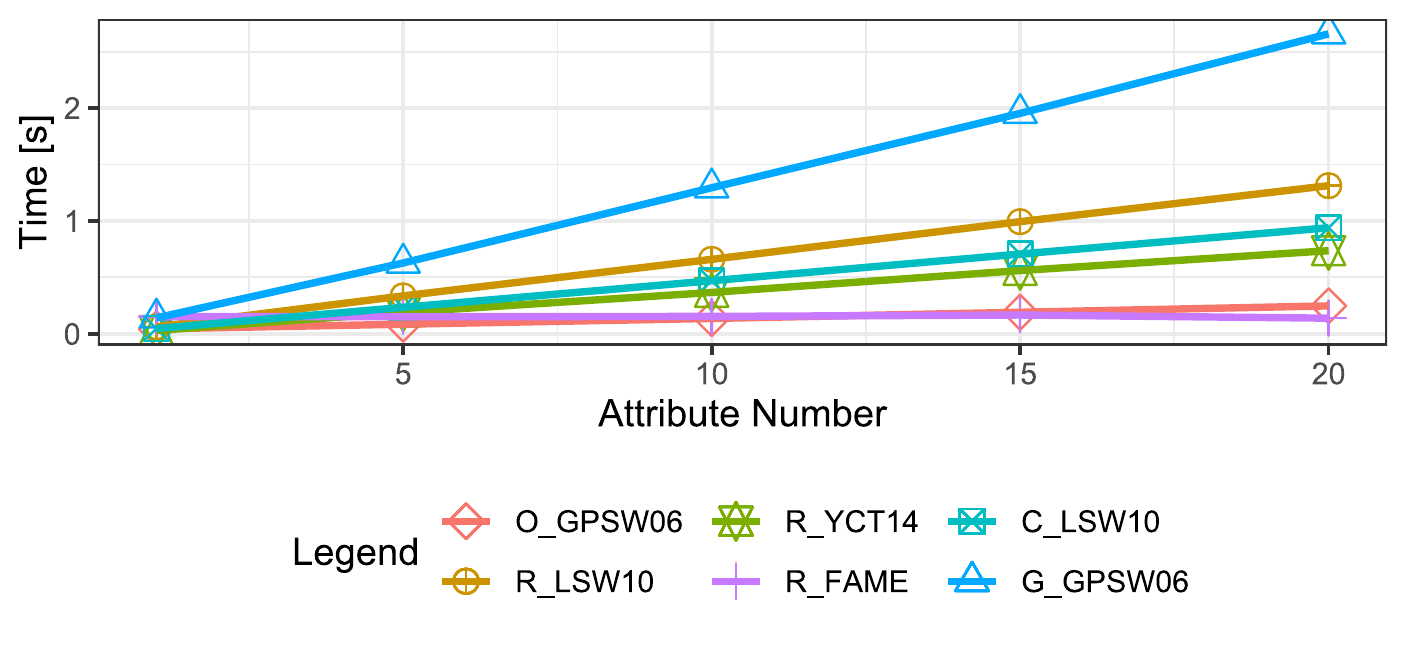}
    \caption{Results in a \ac{RPI4}}
    \label{fig:KPABE_Dec-RPI4}
\end{subfigure}
\hfill
\begin{subfigure}{1\columnwidth}
    \includegraphics[width=1\textwidth]{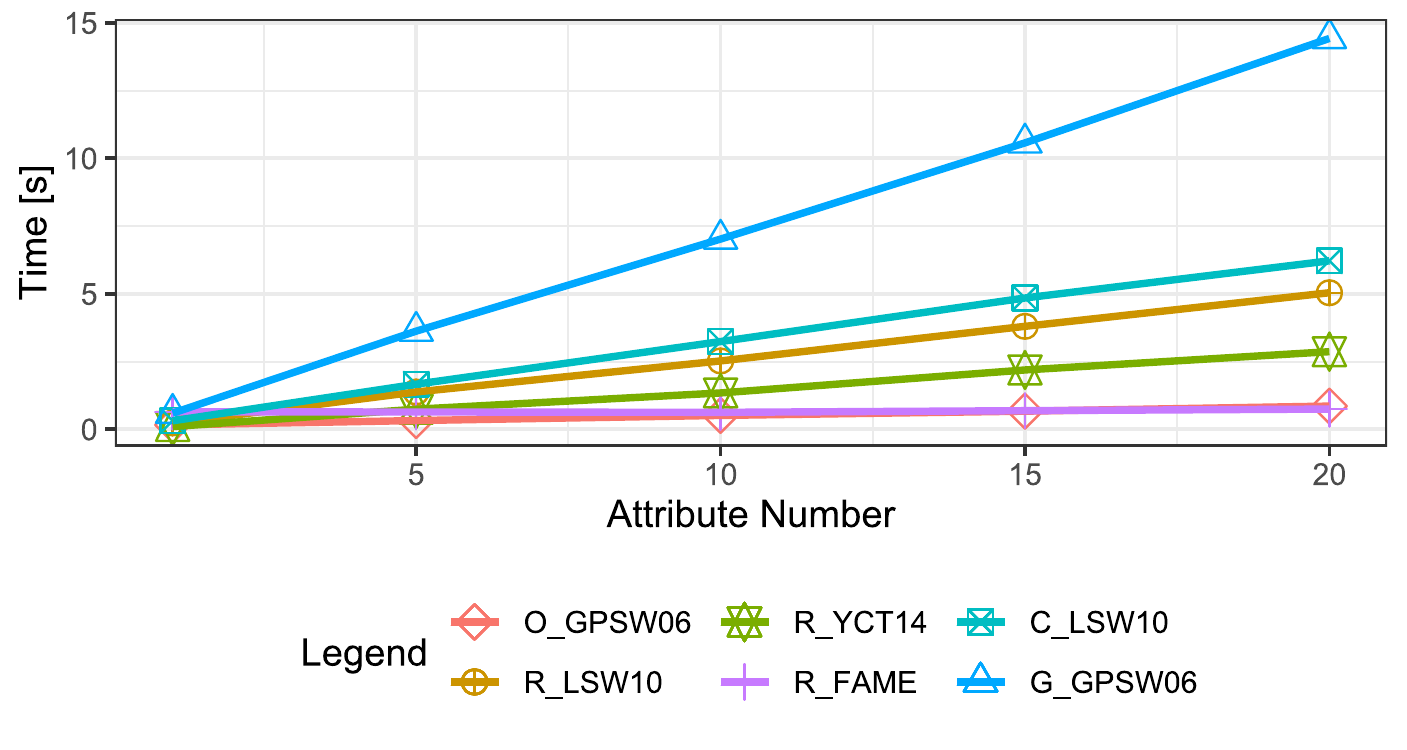}
    \caption{Results in a \ac{RPI0}}
    \label{fig:KPABE_Dec-RPI0}
\end{subfigure}
\caption{KP-ABE decryption times.}
\label{fig:KPABE_Dec}
\end{figure}

\added{Finally, decryption times for KP-ABE are depicted in Figure \ref{fig:KPABE_Dec}. It can be seen that KP-ABE FAME has constant decryption time, as with its CP-ABE version. However, the only library implementing it is\textit{Rabe} (R\_FAME).}


\added{As a result of the FAME's constant decryption times, the fastest decryption scheme depends on the number of attributes of the \ac{CT}. Therefore, O\_GPSW06 is the fastest scheme for less than 11 attributes in \ac{RPI4} and less than 15 attributes in \ac{RPI0}. \ac{RPI4} takes 136 ms to decrypt a \ac{CT} with 10 attributes using O\_GPSW06; and for the same scheme, but with 15 attributes, \ac{RPI0} takes 670 ms. Meanwhile, for more than 11 attributes, R\_FAME becomes the fastest option in \ac{RPI4}, taking 150ms; and for more than 15 in \ac{RPI0}, taking 700ms. Meanwhile, the slowest scheme is G\_GPSW06.}


\subsection{Decentralized CP-ABE} \label{Sec:6.3}

\begin{figure}[!h] 
\centering
\begin{subfigure}{1\columnwidth}
    \includegraphics[width=1\textwidth]{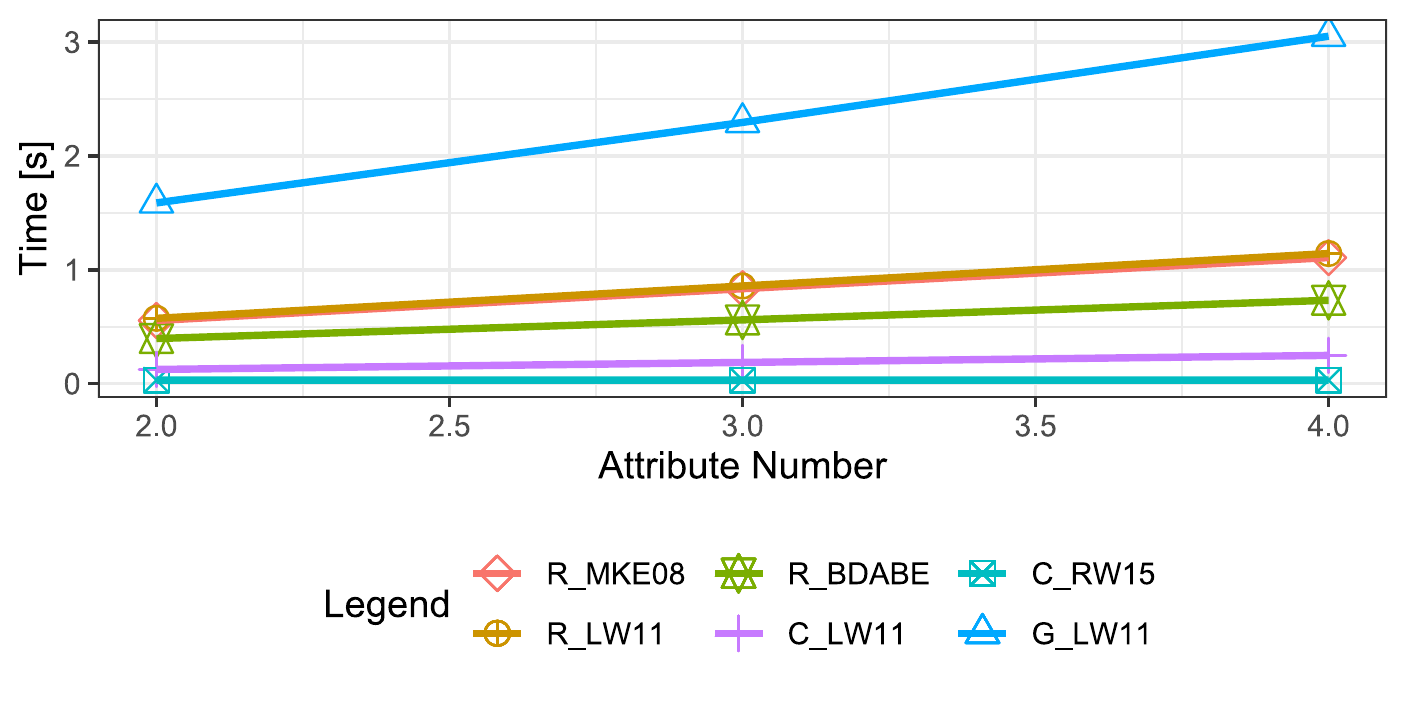}
    \caption{Results in a \ac{RPI4}}
    \label{fig:dCPABE_AuthSetup-RPI4}
\end{subfigure}
\hfill
\begin{subfigure}{1\columnwidth}
    \includegraphics[width=1\textwidth]{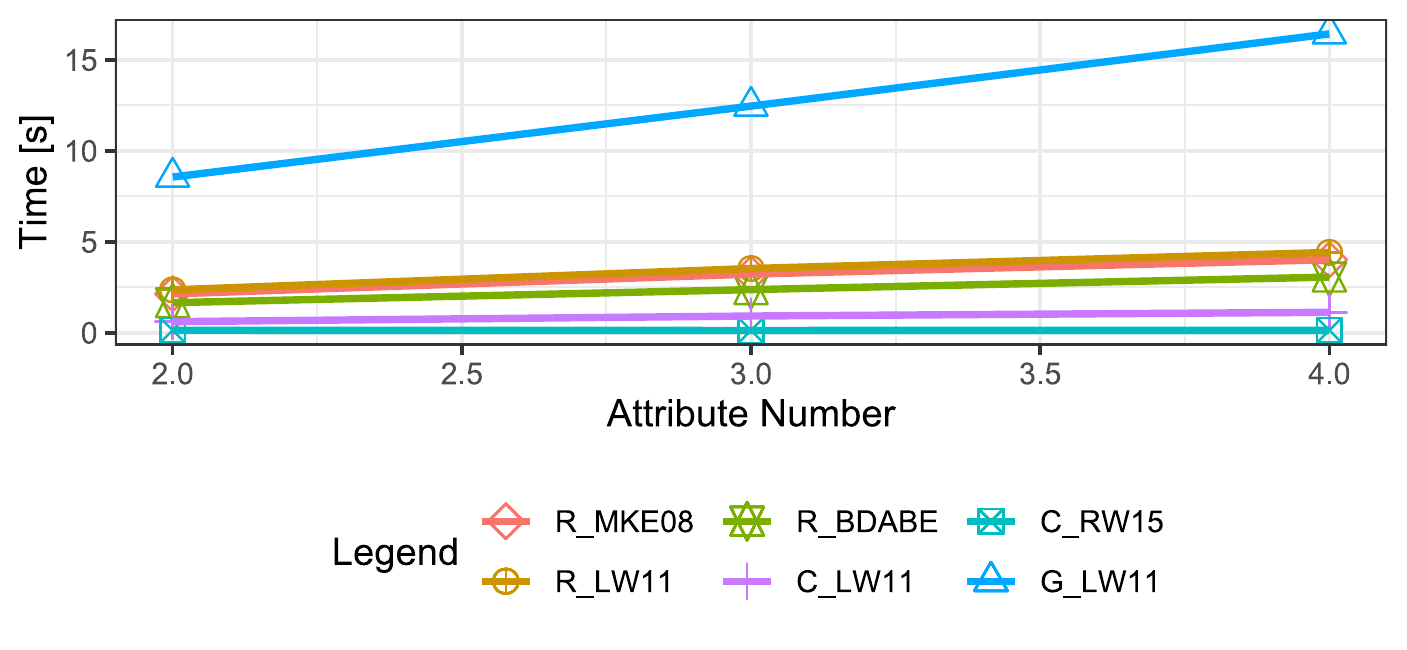}
    \caption{Results in a \ac{RPI0}}
    \label{fig:dCPABE_AuthSetup-RPI0}
\end{subfigure}
\caption{dCP-ABE Authority Setup times.}
\label{fig:dCPABE_AuthSetup}
\end{figure}

\added{Decentralized CP-ABE has an added operation to those shown for KP-ABE or CP-ABE: authority setup. Although every scheme requires a setup, decentralized schemes require the initialization of each authority. Some decentralized schemes allow users to setup authorities at any point in the system's lifetime. Thus, it is interesting to quantify how much time these setups can take. The times can be visualized in Figure \ref{fig:dCPABE_AuthSetup}.}

\added{As Figure \ref{fig:dCPABE_AuthSetup} presents, authorities setup times vary widely between schemes. The figure shows the time it takes to configure five authorities. Specifically, it shows the time evolution depending on how many attributes each manages. C\_RW15 is the fastest one in both devices, taking 30ms for the five authorities to be set up in a \ac{RPI4}, with each of them controlling four attributes each. The \ac{RPI0} requires 132ms for the same operation. The closest scheme in execution performance to C\_RW15 is C\_LW11. However, an analysis of the results shows that C\_LW11 takes 246.84ms in RPI4, presenting a 157\% difference compared to C\_RW15. The same ratio is maintained at RPI0, where C\_LW11 takes 1.12s. Therefore, even the second fastest scheme is no match for C\_RW15. It is noteworthy that RW15 is the evolution of LW11, and thus stands to reason that RW15 is more efficient in some of its operations. Meanwhile, G\_LW11 is the slowest, maintaining \textit{GoFE} as the slowest library for many schemes and operations.}

\begin{figure}[!h] 
\centering
\begin{subfigure}{1\columnwidth}
    \includegraphics[width=1\textwidth]{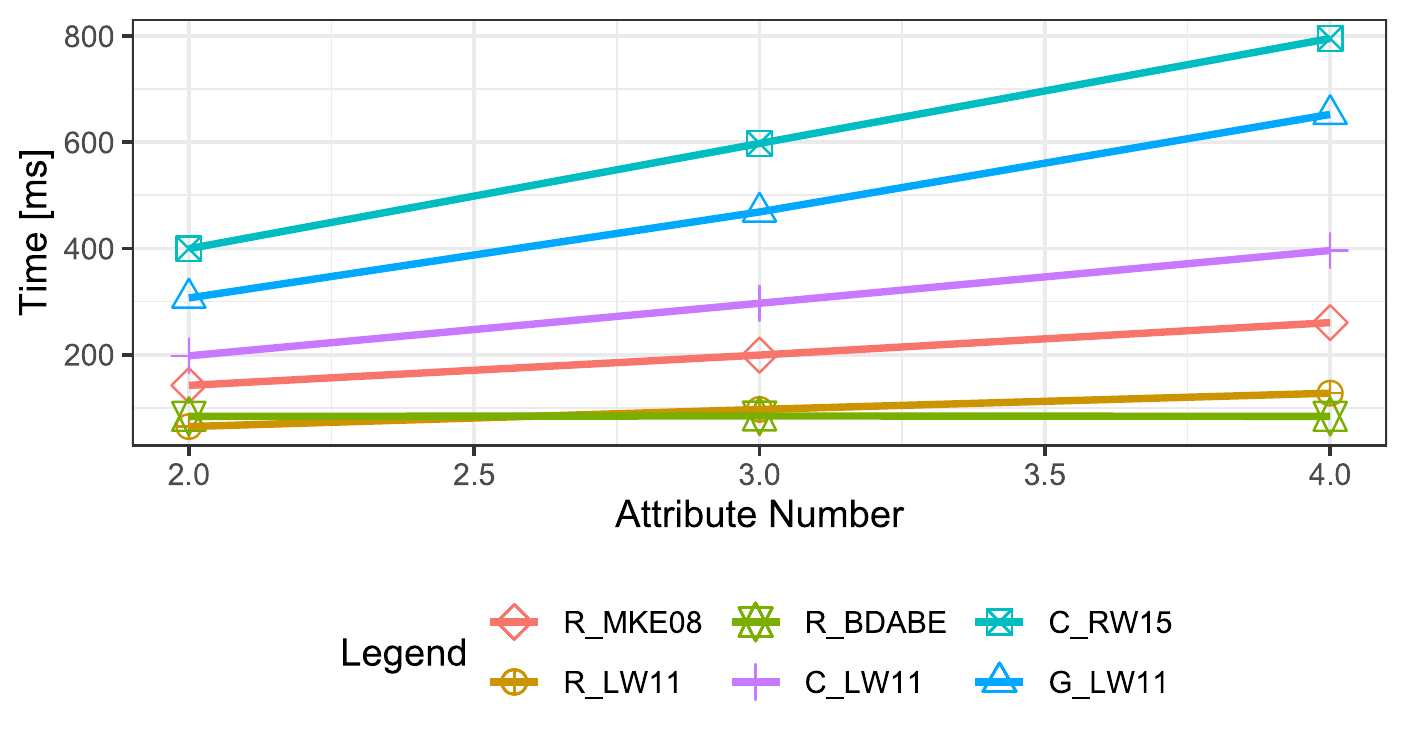}
    \caption{Results in a \ac{RPI4}}
    \label{fig:dCPABE_KeyGen-RPI4}
\end{subfigure}
\hfill
\begin{subfigure}{1\columnwidth}
    \includegraphics[width=1\textwidth]{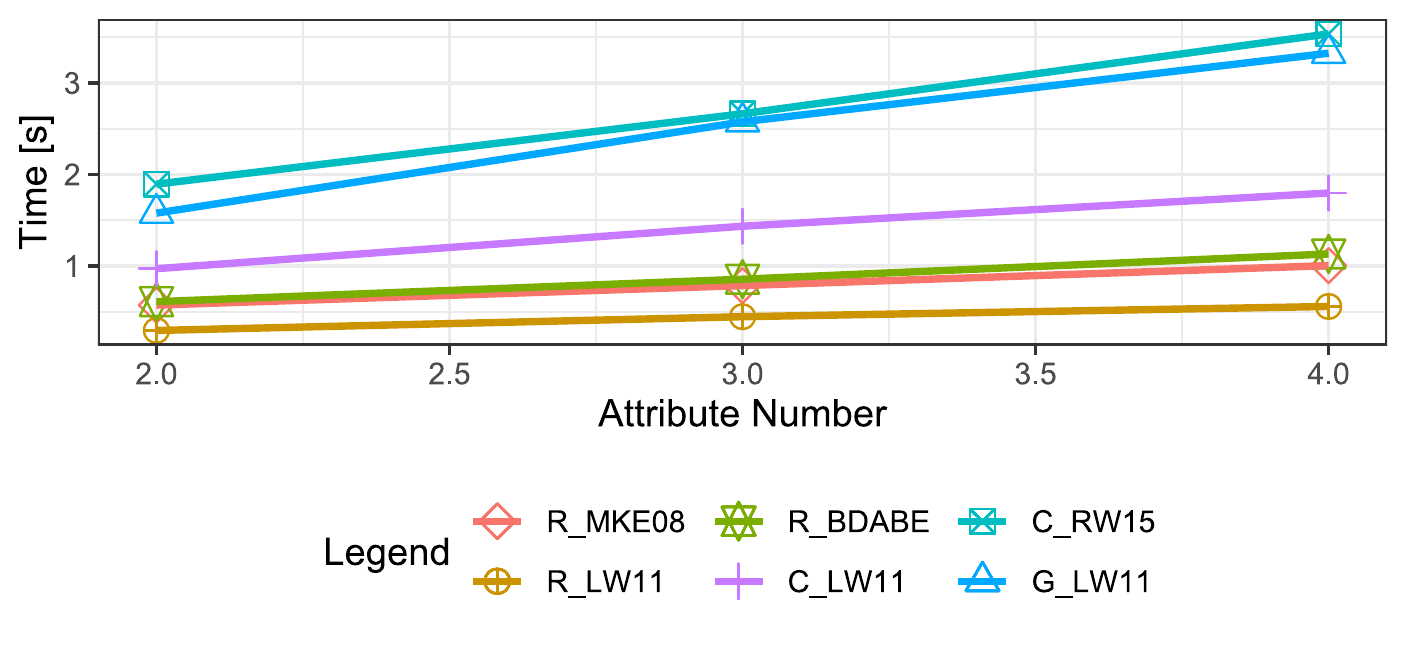}
    \caption{Results in a \ac{RPI0}}
    \label{fig:dCPABE_KeyGen-RPI0}
\end{subfigure}
\caption{dCP-ABE Key Generation times.}
\label{fig:dCPABE_KeyGen}
\end{figure}

\added{After the authorities are setup, they can start generating users' secret keys. For this experiment, we work with five authorities, each of which controls a variable amount of attributes, from 2 to 4. The result of key generation times can be visualized in Figure \ref{fig:dCPABE_KeyGen}.}

\added{As shown in Figure \ref{fig:dCPABE_KeyGen}, the fastest \ac{dCP-ABE} scheme for key generation in the \ac{RPI0} is R\_LW11, which takes 557ms for the case of requesting 4 attributes from each AA. In contrast, on \ac{RPI4}, the fastest scheme for 3 attributes or more is R\_BDABE. For the case of asking 4 attributes to each AA, R\_BDABE takes 33\% less than R\_LW11 (85.25ms in R\_BDABE VS 127ms in R\_LW11). In the case of 3 attributes or less, the fastest scheme is R\_LW11. However, the advantage presented by R\_BDABE is so marginal that when the processing capabilities are reduced (case of \ac{RPI0}), this advantage disappears. Moreover, as introduced during the qualitative evaluation in Section \ref{Sec:4}, to the result of R\_BDABE, we have to add the time needed to interact with the Blockchain, which can be long and unpredictable. Thus, even for \ac{RPI4}, R\_LW11 can be considered the fastest key generation scheme, taking 127ms for the worst case.}

\begin{figure}[!ht] 
\centering
\begin{subfigure}{1\columnwidth}
    \includegraphics[width=1\textwidth]{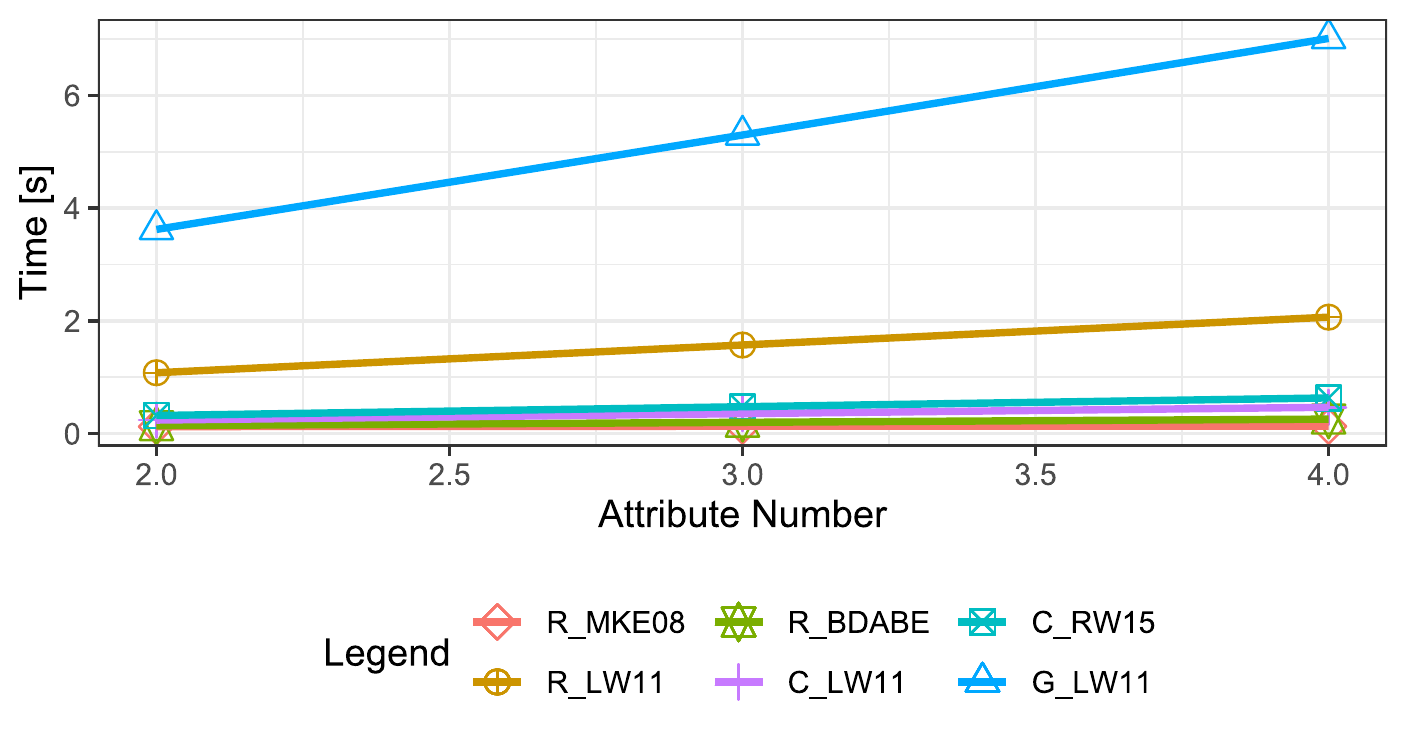}
    \caption{Results in a \ac{RPI4}}
    \label{fig:dCPABE_EncComplete-RPI4}
\end{subfigure}
\hfill
\begin{subfigure}{1\columnwidth}
    \includegraphics[width=1\textwidth]{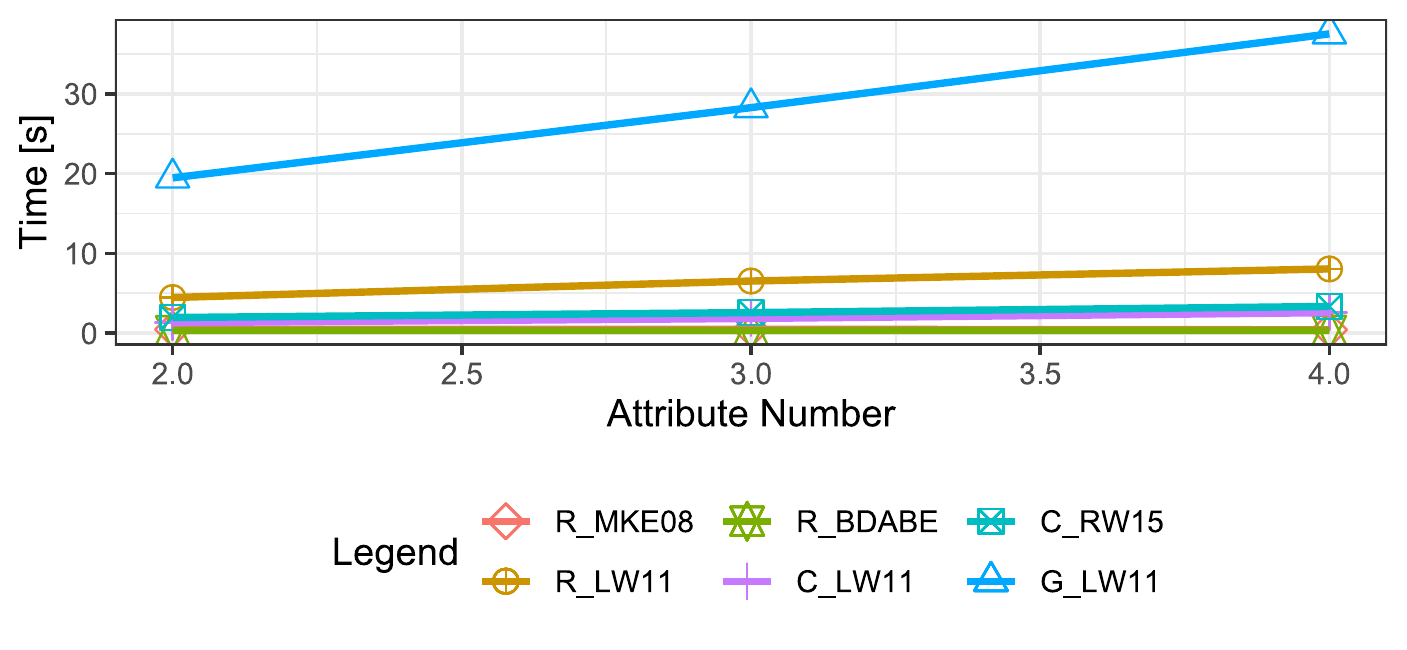}
    \caption{Results in a \ac{RPI0}}
    \label{fig:dCPABE_EncComplete-RPI0}
\end{subfigure}
\caption{dCP-ABE encryption times.}
\label{fig:dCPABE_EncComplete}
\end{figure}

\added{Finally, the slowest scheme is C\_RW15 in both cases. However, the difference between \ac{RPI0} and \ac{RPI4} is noteworthy. While G\_LW11 is the slowest scheme in \ac{RPI4} with a noticeable difference from C\_RW15, this difference is much smaller in \ac{RPI0}. Moreover, in the case of having to request 4 attributes, G\_LW11 on RPI0 takes 3.32s while C\_RW15 takes 3.53s. The little difference between the two schemes is related to the implementation of \textit{GoFE}. In all the results, it has been observed how this library, when implemented in \ac{RPI0}, achieves much worse results than in \ac{RPI4}. However, key generation is an operation that takes place only when users require a key, and its impact is limited. Meanwhile, encryption and decryption take place more often. Encryption results are presented in Figure \ref{fig:dCPABE_EncComplete}. } 

\begin{figure}[!h] 
\centering
\begin{subfigure}{1\columnwidth}
    \includegraphics[width=1\textwidth]{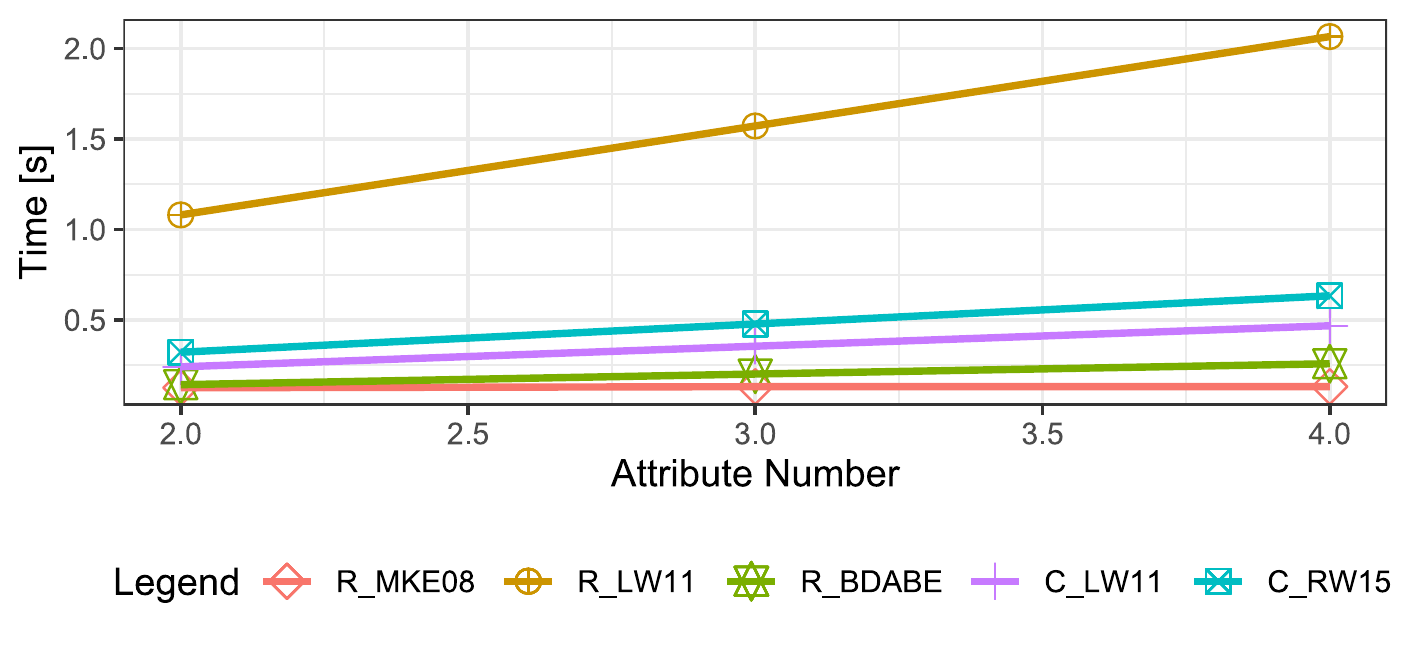}
    \caption{Results in a \ac{RPI4}}
    \label{fig:dCPABE_Enc-RPI4}
\end{subfigure}
\hfill
\begin{subfigure}{1\columnwidth}
    \includegraphics[width=1\textwidth]{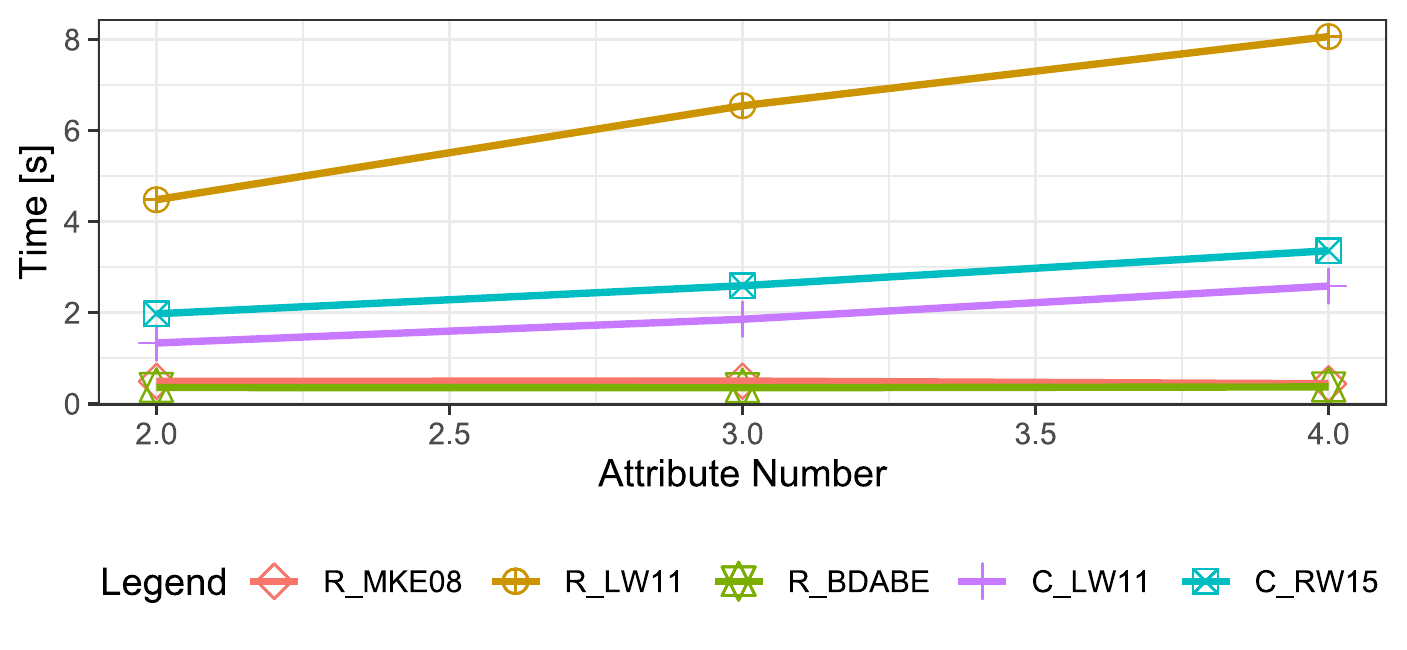}
    \caption{Results in a \ac{RPI0}}
    \label{fig:dCPABE_Enc-RPI0}
\end{subfigure}
\caption{dCP-ABE encryption times.}
\label{fig:dCPABE_Enc}
\end{figure}

\added{Figure \ref{fig:dCPABE_EncComplete} shows that the slower encryption scheme is G\_LW11, taking up to 5s for a policy of 20 attributes (distributed homogeneously among five authorities) in a \ac{RPI4}. For the same operation, the \ac{RPI0} requires 37.5s. In fact, LW11 is generally one of the slowest schemes since it is also the slowest dCP-ABE scheme for\textit{Rabe} (R\_LW11), which takes 1.3 seconds for the worst case in the \ac{RPI4} and 8s for a \ac{RPI0}. To better see the results of the rest of the schemes, we provide Figure \ref{fig:dCPABE_Enc}.}

\added{Figure \ref{fig:dCPABE_Enc} allows us to better see the encryption results of the fastest schemes. It shows that the fastest one is R\_MKE08, taking 131ms for encryption in \ac{RPI4}. This same operation in the \ac{RPI0} takes 439ms. Although R\_BDABE is almost as fast as R\_MKE08, one should consider the time delays related to the Blockchain operations.}

\begin{figure}[!h] 
\centering
\begin{subfigure}{1\columnwidth}
    \includegraphics[width=1\textwidth]{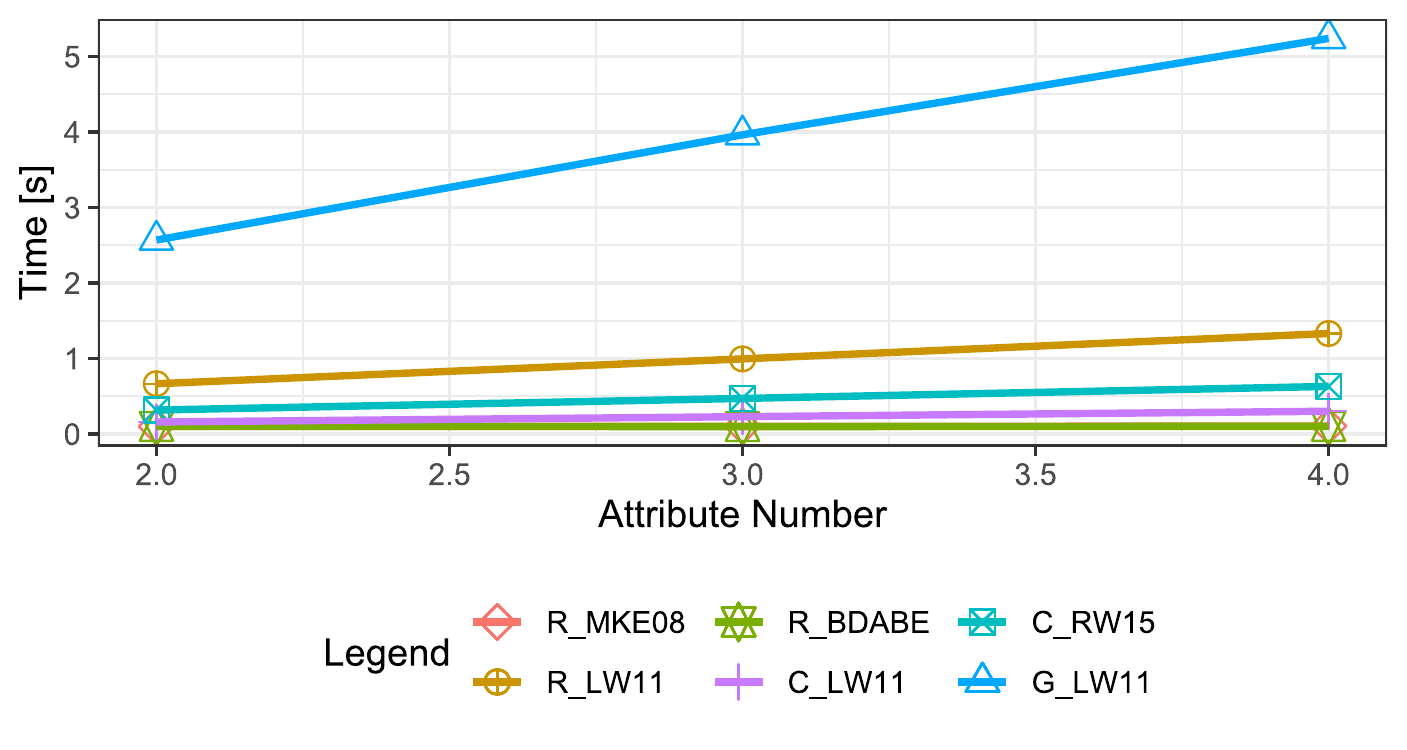}
    \caption{Results in a \ac{RPI4}}
    \label{fig:dCPABE_DecComplete-RPI4}
\end{subfigure}
\hfill
\begin{subfigure}{1\columnwidth}
    \includegraphics[width=1\textwidth]{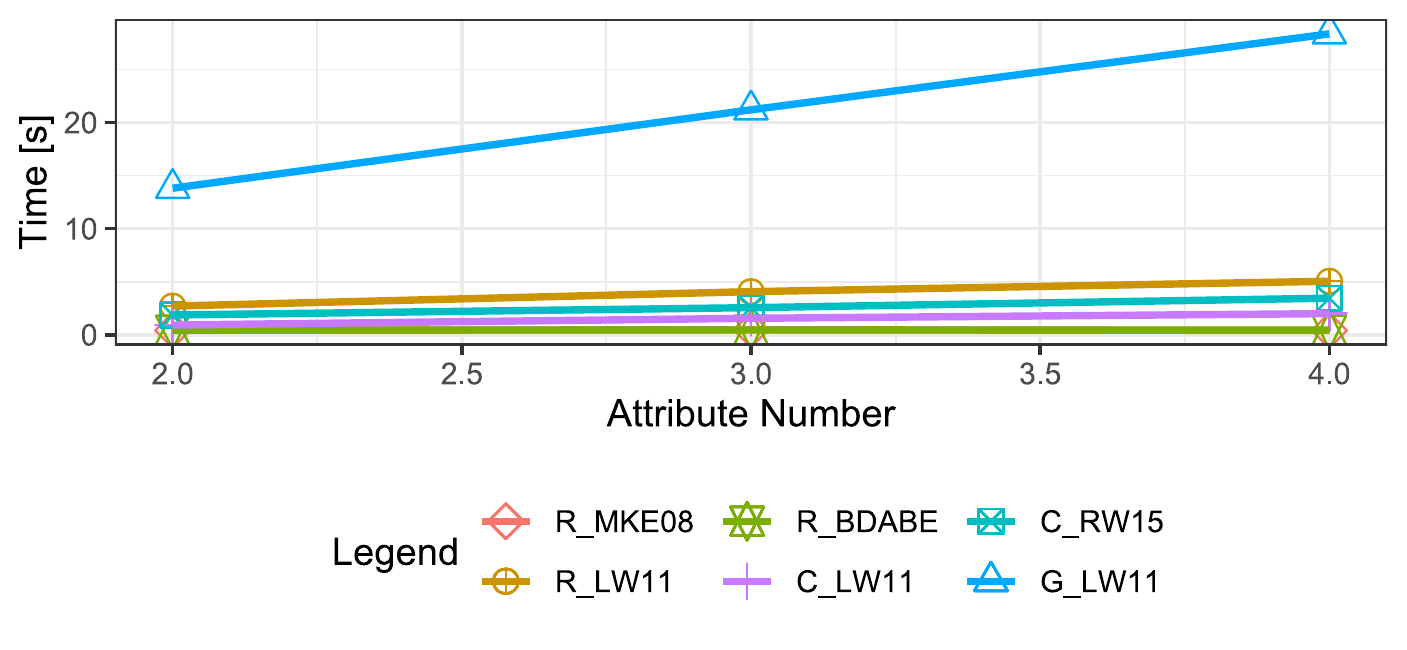}
    \caption{Results in a \ac{RPI0}}
    \label{fig:dCPABE_DecComplete-RPI0}
\end{subfigure}
\caption{dCP-ABE decryption times.}
\label{fig:dCPABE_DecComplete}
\end{figure}

\added{Finally, after encrypting information, we depict decryption time results in Figure \ref{fig:dCPABE_DecComplete}. We see that results are somewhat analogous to those of encryption. Decryption is faster than encryption, but the schemes' performance is equivalent. Once again, the results show that G\_LW11 is the slowest scheme. }

\added{Finally, R\_BDABE and R\_MKW08 are the fastest schemes for decryption. However, as previously stated, BDABE will have an added delay related to the Blockchain when implemented. Thus, the fastest scheme is R\_MKE08, which takes 105ms on average for decryption in a \ac{RPI4} and 510ms for \ac{RPI0}.}

%% file: Docs/7.Discussion.tex
\section{Discussion} \label{Sec:7}

\begin{table*}[tb]
    \renewcommand{\arraystretch}{1.4}
    \caption{Fastest schemes for each considered functions} \label{tab:5}
    \begin{center}
    \resizebox{1\textwidth}{!}{
    \begin{tabular}{ll|c:c|c:c|c:c|c:c}
    	\\ \hhline{~~========}

    	  & 
          & \multicolumn{2}{c|}{Auth Setup}
          & \multicolumn{2}{c|}{KeyGen}
          & \multicolumn{2}{c|}{Enc.} 
    	  & \multicolumn{2}{c}{Dec.}
    	  \\ \cline{3-10}
    	  
    	  & & \ac{RPI0} & \ac{RPI4} & \ac{RPI0} & \ac{RPI4} & \ac{RPI0} & \ac{RPI4} & \ac{RPI0} & \ac{RPI4}  
    	  \\ \hline \hline
    	\multicolumn{1}{c|}{\multirow{3}{*}{\rothead{Scheme Type}}} & CP-ABE & --- & --- & O\_W11 & O\_W11 & O\_W11 & O\_W11 & C\_BSW07 & C\_BSW07 
    		\\ \cline{2-10}
    		\multicolumn{1}{c|}{} & KP-ABE & --- & --- & R\_YCT14 & R\_YCT14 & O\_GPSW06 & O\_GPSW06 & \begin{tabular}{cc} O\_GPSW06 (\textless 15 att)\\ R\_FAME (\textgreater 15 att)\end{tabular} & \begin{tabular}{cc} O\_GPSW06 (\textless 10 att)\\ R\_FAME (\textgreater 10 att)\end{tabular}
    			\\ \cline{2-10}
    		\multicolumn{1}{c|}{} & dCP-ABE & C\_RW15 & C\_RW15 & R\_LW11 & R\_LW11 & R\_MKE08 & R\_MKE08 & R\_MKE08 & R\_MKE08
    			\\ \hline \hline
    	\multicolumn{1}{c|}{\multirow{4}{*}{\rothead{Language}}} & Python & RW15 & RW15 & \begin{tabular}{cc}BSW07(\textless 5 att) \\ W11 (\textgreater 5 att)\end{tabular} & \begin{tabular}{cc}BSW07(\textless 10 att) \\ W11(\textgreater 10 att)\end{tabular} & LSW10 & LSW10 & BSW07 & BSW07 
    		\\ \cline{2-10}
    		\multicolumn{1}{c|}{} & Golang & LW11 & LW11 & GPSW06 & GPSW06 & GPSW06 & GPSW06 & \begin{tabular}{cc}FAME (\textgreater 6 att) \\ GPSW06 (\textless 6 att)\end{tabular} & \begin{tabular}{cc}FAME (\textgreater 6 att) \\ GPSW06 (\textless 6 att)\end{tabular}
    			\\ \cline{2-10}
    		\multicolumn{1}{c|}{} & C++ 	 & --- & --- & W11 & W11 & GPSW06 & GPSW06 & GPSW06 & GPSW06 
    			\\ \cline{2-10}
    		\multicolumn{1}{c|}{} & Rust   & MKE08  &  MKE08 & YCT14 & YCT14 & \begin{tabular}{cc}MKE08 (\textgreater 5 att) \\ YCT14 (\textless 5 att)\end{tabular} & \begin{tabular}{cc}MKE08 (\textgreater 5 att) \\ YCT14 (\textless 5 att)\end{tabular}  & MKE08 & MKE08
    			\\ \hline \hline
    \end{tabular}
    }
    \end{center}
\end{table*}

\added{As a result of our experimental evaluations, we provide Table \ref{tab:5}, which summarizes the most efficient schemes for each case.The performance of BDABE relies on the type of Blockchain used, and developers must be aware of this when evaluating the results in this section.}

\added{Table \ref{tab:5} presents the results according to the type of ABE (\ac{CP-ABE}, \ac{KP-ABE}, and \ac{dCP-ABE}) and the implementation language. Sometimes developers may require a specific ABE type but without a specific scheme. Table \ref{tab:5} shows which scheme from which library is the most efficient. Similarly, we also present results tied to a specific library, as developers may be constrained to specific languages.}

\added{Regarding the scheme type, for developers working with \ac{CP-ABE}, the fastest scheme for key generation is O\_W11. One of the main advantages of this scheme is that it is also the fastest for encryption. However, developers should consider that \ac{CP-ABE} schemes usually have slower decryption time than encryption, and O\_W11 is no exception. In fact, if developers need a fast decryption \ac{CP-ABE} scheme, they should choose C\_BSW07. This scheme has a good balance between the three operations, since it is the second fastest for key generation and encryption in both devices.}

\added{When dealing with \ac{KP-ABE}, the fastest scheme for key generation is R\_YCT14. Key generation only takes place when system users request them, but fast generation is useful in systems where new users are continually added. Meanwhile, the fastest scheme for encryption is O\_GPSW06. One of the advantages of this scheme is that it is also the second fastest for key generation. In addition, O\_GPSW06 is also the fastest scheme for \ac{AP}s with less than 10 attributes in \ac{RPI4} and less than 15 attributes in \ac{RPI0}. However, if the system has \ac{AP}s with more attributes, R\_FAME becomes a better choice. R\_FAME has constant decryption times, so its decryption time is independent of the complexity of the \ac{AP}s.}

\added{Regarding \ac{dCP-ABE}, developers now have to consider the time required to set up authorities. Some schemes only allow developers to do this during the system setup, but others allow AAs to be set up throughout the system's lifetime. In this regard, the fastest scheme is C\_RW15. However, it is also one of the slowest schemes for key generation, encryption, and decryption. For faster key generation, the best option is R\_LW11. However, this scheme is also the second slowest for encryption and the slowest for decryption. Thus, the fast key generation is hardly compensated by the rest of the delays since encryption and decryption are more common tasks. However, as mentioned earlier, systems with many new users may benefit from this. Finally, for fast encryption and decryption, the best option is R\_MKE08. It is noteworthy that this scheme is also the second fastest one for key generation.}

\added{Table \ref{tab:5} also summarizes the fastest schemes for each library. Not all libraries implement the same schemes, so the options for some may be reduced. For example, \textit{OpenABE} only provides one \ac{CP-ABE} and one \ac{KP-ABE} scheme. However, we can see that W11 is the fastest scheme for setupkey generation in both \textit{OpenABE} and Charm, and GPSW06 is the fastest for encryption in \textit{GoFE} and \textit{OpenABE}. In the case of textit{Rabe}, we can see how MKE08 is one of the most balanced schemes: it is the fastest scheme provided by textit{Rabe} for authority generation, encryption, and decryption.}

\added{Finally, other results that might be useful for developers and are not contemplated in \ref{tab:5} may be:}

\begin{itemize}
    \item \added{BDABE is the only Blockchain-based scheme, and textit{Rabe} is the only library implementing it.}
    \item \added{\textit{OpenABE} was not originally designed for ARM architectures. However, it can be adapted to ARMv7 and ARMv8 architectures}\footnote{https://github.com/IBM/openabe}, \added{and with further modifications}\footnote{https://github.com/relic-toolkit/relic/issues/211} \added{to ARMv6. }
    \item \added{Some users may require schemes with attribute revocation. This requirement is out of the scope of the survey presented in this paper. However, the schemes that claim to provide this feature are LSW10, TimePRE, DAC-MACS, YJ14, and BDABE.}
    \item \added{YCT14, YJ14, and DAC-MACS are broken, and therefore should be avoided.}
\end{itemize}

%% file: Docs/8.Conclusions.tex
\section{Conclusions} \label{Sec:8}
This paper provides a qualitative and quantitative comparison of 11 ABE libraries. We first highlight those implementing vulnerable \replaced{ABE}{CP-ABE} schemes and identify the mathematical library underneath. This, among other criteria, \replaced{has discouraged using several schemes.}{has been used to discourage using several libraries.}

\added{We also provide a qualitative evaluation of each scheme provided by the libraries and their distinctive features. Since all schemes are implemented in hybrid mode, we also identify and analyze the AES mode used by each library.}

\added{We select the four best candidates and quantitatively assess their efficiency based on the qualitative evaluation. We measure efficiency in terms of how much time each scheme takes to perform basic operations like key generation, encryption, and decryption.}

\replaced{As a result of our experimental evaluations, we provide a table in which we indicate which schemes are faster based on their approach, programming language, and the device running them.}{As a result of our experimental evaluations, we conclude that \textit{OpenABE} has the fastest key generation and encryption; and also decryption if less than five attributes are considered. Meanwhile, if developers require a larger number of attributes and a reduced time for decryption, Charm or textit{Rabe} are recommended, although this comes at the expense of additional key generation and encryption times. }

\added{Developers may require efficiency at different points in the process: key generation, authority setup, encryption, or decryption. There is no one-size-fits-all scheme, because no one scheme can be efficient in all aspects of the process. Thus, developers should analyze which step will require lower computational time. Our discussion provides developers with the tools to support them in selecting the most suitable library and scheme.}



%% file: Docs/ACKs.tex
\section*{Acknowledgment}
The European commission financially supported this work through Horizon Europe program under the IDUNN project (grant agreement N$^{\circ}$ 101021911). It was also partially supported by the \textit{Ayudas Cervera para Centros Tecnológicos} grant of the Spanish Centre for the Development of Industrial Technology (CDTI) under the project EGIDA (CER-20191012), and by the Basque Country Government under the ELKARTEK program, project REMEDY - REal tiME control and embeddeD securitY (KK-2021/00091).